\documentclass[11pt]{article}

\usepackage[utf8]{inputenc}
\usepackage[T1]{fontenc}
\usepackage{hyperref}
\usepackage{url}
\usepackage{booktabs}
\usepackage{amsfonts}
\usepackage{amsmath}
\usepackage{amssymb}
\usepackage{csquotes}
\usepackage{nicefrac}
\usepackage{microtype}
\usepackage{graphicx}
\usepackage{fancyhdr}
\usepackage{natbib}
\usepackage[margin=0.7in, headheight=14pt]{geometry}
\usepackage{algorithm}
\usepackage{algpseudocode}
\usepackage{enumitem}
\usepackage{listings}
\usepackage{xcolor}
\usepackage{framed}
\usepackage{tikz}
\usepackage{algpseudocode}
\usepackage{geometry}
\usepackage{authblk}

\definecolor{codebg}{RGB}{248,248,248}
\definecolor{codeframe}{RGB}{220,220,220}
\definecolor{codecomment}{RGB}{0,128,0}
\definecolor{codekeyword}{RGB}{0,0,150}
\definecolor{codestring}{RGB}{163,21,21}

\lstdefinestyle{compactpython}{
    language=Python,
    backgroundcolor=\color{codebg},
    basicstyle=\ttfamily\small,
    keywordstyle=\color{codekeyword}\bfseries,
    stringstyle=\color{codestring},
    commentstyle=\color{codecomment}\itshape,
    frame=single,
    rulecolor=\color{codeframe},
    framerule=0.5pt,
    columns=fullflexible,
    keepspaces=true,
    showstringspaces=false,
    breaklines=true,
    breakatwhitespace=true,
    tabsize=4,
    xleftmargin=6pt,
    xrightmargin=6pt,
    aboveskip=0.6em,
    belowskip=0.6em,
    framesep=4pt
}

\usetikzlibrary{arrows.meta, positioning, shapes.geometric}

\definecolor{airptshade}{gray}{0.95}
\definecolor{mainresultbg}{RGB}{236,243,252}
\definecolor{mainresultborder}{RGB}{170,196,230}
\newenvironment{airptbox}{%
    \MakeFramed{\advance\hsize-\width \FrameRestore}%
    \noindent\textbf{Human-Edited AI-Generated Summary.} The following algorithm explanations are based on the AI-generated run report produced by \textsc{MadEvolve}. They were condensed and edited by the authors to improve readability. \par\medskip
}{%
    \endMakeFramed
}

\title{MadEvolve: Evolutionary Optimization of Trading Systems with Large Language Models}

\author[1,*]{Yurii Kvasiuk}
\author[1,*]{Tianyi Li}
\author[2]{Owen Colegrove}
\author[1]{Moritz M\"unchmeyer}

\affil[1]{Department of Physics, University of Wisconsin--Madison}
\affil[2]{Event Horizon Labs}
\affil[*]{Corresponding authors: \texttt{kvasiuk@wisc.edu}, \texttt{tianyi.li@wisc.edu}}

\date{}

\begin{document}
\maketitle
\begin{tikzpicture}[remember picture, overlay]
    \node[anchor=south east, xshift=-0.7in, yshift=0.7in] at (current page.south east)
        {\includegraphics[width=1in]{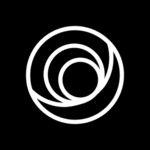}};
\end{tikzpicture}

\begin{abstract}
We explore the application of LLM-driven algorithm optimization to several common tasks in quantitative finance. MadEvolve, a general-purpose algorithm optimization framework inspired by DeepMind's Alpha-Evolve, was recently developed to optimize algorithms in computational cosmology. Here we demonstrate the utility of MadEvolve to optimize algorithmic trading strategies and alpha generation at the example of Bitcoin trading. On our simulation and backtesting setup, we achieve significant improvements on all tasks we considered, such as evolving feature sets for signal generation, optimizing separate components of the trading strategy, and jointly evolving the feature pipeline together with the execution strategy. Additionally, we compare our method to other agentic search approaches, specifically Claude Code, and carefully evaluate p-hacking probabilities on our simulation setup. Our findings strongly support the utility of AI-driven agentic and evolutionary algorithms for algorithmic trading and quantitative finance.
\end{abstract}

\clearpage

\setcounter{tocdepth}{2}
\tableofcontents
\clearpage

\section{Introduction}
\label{sec:intro}

The rate of improvement in artificial intelligence capabilities is unprecedented. Model performance on reasoning benchmarks has improved by orders of magnitude in the span of two years \citep{openai2024o1, anthropic2024claude3, google2024gemini}, and the cost of inference has dropped correspondingly. This trajectory has shifted the question from \emph{whether} AI can contribute to complex intellectual tasks to \emph{how quickly} it will reshape them.

A particularly powerful emerging pattern is the \emph{agentic loop}: an LLM-driven system that iteratively proposes solutions, evaluates them against a reward signal, and refines its approach based on feedback. Google DeepMind's AlphaEvolve \citep{alphaevolve2025} demonstrated that such loops can discover novel algorithms for well-defined mathematical and computational problems, matching or exceeding decades of human research. The key insight is that when a problem admits a clear, automated evaluation function, the search process can run autonomously at scale, exploring vastly more of the solution space than any human team could.

This raises a fundamental question: \emph{Can agentic loops optimize problems with noisy reward functions?} AlphaEvolve's targets---matrix multiplication, bin packing, compiler heuristics---have deterministic or near-deterministic evaluation. But most real-world optimization problems do not. The reward signal is stochastic, the evaluation is expensive, and the risk of overfitting to noise is substantial. Understanding where agentic loops break down as reward functions become noisier is among the most important open questions in applied AI.

Financial markets represent a prime example of noisy optimization: a domain where automation could unlock enormous value, but where the reward signal actively resists naive search. Markets generate abundant quantitative feedback---profit and loss (PnL), execution costs, $R^2$ of forecasts against realized returns---that could in principle serve as fitness functions. But these signals are non-stationary, confounded by regime changes, and reflexive (deploying a successful strategy erodes its own edge). Worse, the large dimensionality and historical richness of financial data make it trivially easy to ``discover'' patterns that are pure artifacts of multiple testing---the well-documented problem of backtest overfitting \citep{bailey2014backtest}. Any system that evolves trading strategies must therefore answer a prior question: \emph{are we doing research, or are we p-hacking?}

We study this problem concretely by applying agentic loops to several trading system components of increasing difficulty, and analyzing the results along four axes:

\begin{enumerate}[label=(\roman*)]
    \item \textbf{Iteration depth}: Does continued evolution improve out-of-sample results, or does it merely overfit? In Sec. \ref{sec:discussion} we compare observed IS--OOS degradation against the theoretical discount predicted by multiple-testing corrections \citep{bailey2014backtest} to assess whether the loop is discovering genuine structure or p-hacking. We further investigate to what extent overfitting depends on iteration depth (see e.g. Fig. \ref{fig:run5_is_oos_degradation}).
    \item \textbf{Model quality}: How do different frontier models contribute to the evolutionary process? Our ensemble of five LLMs exhibits striking variation in mutation quality across problem types, with improvement rates ranging from 3\% to 64\% depending on model and task. We analyze these patterns in Section~\ref{sec:model_analysis}.
    \item \textbf{Problem complexity}: Does performance degrade gracefully as the optimization target becomes more complex, or does it collapse beyond a threshold? Motivated by evidence that LLM reasoning performance exhibits abrupt phase transitions---near-perfect on simple instances but collapsing to near-zero beyond a critical complexity threshold \citep{shojaee2025illusion}---we vary problem complexity from single-component tuning to full-system joint optimization, culminating in a run that jointly evolves the feature-engineering pipeline and the execution strategy together (Section~\ref{sec:exp_joint_calcset_strategy}). We find graceful degradation rather than a cliff: joint feature+strategy evolution achieves the highest out-of-sample Sharpe ratio of any of our experiments, though with the largest validation-to-test PnL retention gap.
    \item \textbf{Noise level}: How does optimization quality degrade as the reward signal becomes noisier? We explore this question by tackling optimization problems with very different signal-to-noise ratio of the fitness function---from the near-deterministic regime of execution optimization to the high-noise regime of return forecasting---to map the frontier of what agentic loops can reliably optimize. A summary of our experiments can be found in Sec. \ref{subsec:summary}. 
\end{enumerate}

We find that our general-purpose agentic optimization loop---\textsc{MadEvolve}, described in Section~\ref{sec:framework}---performs remarkably well on these problems. On the algorithmic subproblems (execution, portfolio construction), the evolved solutions improve substantially over hand-engineered baselines while maintaining out-of-sample validity: the IS--OOS degradation ratios remain well below the p-hacking baseline predicted by classical multiple-testing theory \citep{bailey2014backtest}. The improvements are real, not artifacts of data mining. On forecasting---the noisiest domain---the picture is more nuanced, with some signal persisting OOS, but with the optimized prediction model also showing some degree of overfitting. A complementary set of hyperparameter-calibration experiments (Section~\ref{sec:hyperparam_calibration}) shows that the evolved forecast does carry genuine signal, but its advantage in out-of-sample PnL only emerges once the execution hyperparameters are re-tuned to its scale.

The remainder of the paper is organized as follows. Section~\ref{sec:background} reviews related work on agentic optimization and AI in quantitative finance, and introduces the distinction between forecasting and algorithm-optimization problems that we use to organize our experiments. Section~\ref{sec:framework} summarizes the \textsc{MadEvolve} framework and the adaptations we made for the trading domain. Section~\ref{sec:trading_setup} describes the simulation setup and base strategy. Section~\ref{sec:results} presents the five evolution runs, ranging from tuning a single execution component to jointly evolving features and full execution logic, together with the hyperparameter calibration of the execution strategy (Section~\ref{sec:hyperparam_calibration}) and an analysis of how individual models in our ensemble contribute to the search. Section~\ref{sec:claudecode} compares \textsc{MadEvolve} with a less structured Claude Code agent on the same problems. Section~\ref{sec:discussion} returns to the central question of whether the loop is doing research or p-hacking, and Section~\ref{sec:conclusion} concludes.

\section{Background and Related Work}
\label{sec:background}

\subsection{The Exponential Trajectory of AI}

The capabilities of large language models have followed a trajectory that, by most benchmarks, is super-linear. GPT-4 \citep{openai2023gpt4} demonstrated broad competence across reasoning tasks in early 2023; within 18 months, reasoning-specialized models such as o1 \citep{openai2024o1} and Claude 3.5 \citep{anthropic2024claude3} showed substantial improvements on mathematical and scientific reasoning. By early 2026, frontier models achieve non-trivial scores on FrontierMath \citep{frontiermath2025}---a benchmark of original, research-level mathematics---and Humanity's Last Exam \citep{hle2025}, problems that would challenge domain experts.

More importantly for our purposes, the \emph{agentic} capabilities of these models---their ability to use tools, maintain context over long horizons, and self-correct---have improved dramatically. \citet{metr2025} find that the length of tasks AI agents can complete autonomously is doubling approximately every seven months, enabling the transition from zero-shot inference to the multi-step, hour-scale research workflows required for complex optimization. Modern agents can write, execute, and debug code; query APIs; interpret results; and iterate on solutions with minimal human supervision \citep{wang2024survey_agents}. This creates the substrate for automated research: if an agent can write code, run experiments, interpret results, and propose improvements, then any problem with a programmatic evaluation function becomes a candidate for automated optimization. 

\subsection{Automated Research and AlphaEvolve}

The vision of AI-driven scientific discovery has gained significant traction. FunSearch \citep{romera2024funsearch} demonstrated that LLMs paired with evolutionary search could discover novel solutions to open problems in combinatorics. AlphaEvolve \citep{alphaevolve2025} generalized this approach, showing that an agentic loop combining LLM-based code generation with automated evaluation could optimize algorithms across domains including matrix multiplication, compiler heuristics, and hardware design.

The core loop is conceptually simple:
\begin{enumerate}
    \item \textbf{Propose}: An LLM generates or modifies a candidate solution (typically as code).
    \item \textbf{Evaluate}: The candidate is scored against an automated fitness function.
    \item \textbf{Select}: High-scoring candidates are retained; low-scoring ones are discarded.
    \item \textbf{Evolve}: The LLM generates new candidates informed by the best existing solutions.
\end{enumerate}

The power of this approach lies in its scalability: the loop can run thousands of iterations without human intervention, exploring a vast solution space guided by the LLM's ability to make semantically meaningful modifications rather than random mutations.

Concurrent work has begun applying this paradigm directly to quantitative finance. \citet{yun2025quantevolve} present QuantEvolve, a multi-agent system that uses MAP-Elites with island models and hypothesis-driven LLM mutation to evolve complete trading strategies (as executable Python code) for a small universe of US equities and index futures. Their architecture---LLM-guided code generation, quality-diversity selection, island migration---follows the AlphaEvolve template, with a finance-specific evaluation harness (Zipline backtesting) and a multi-agent decomposition into research, coding, and evaluation roles. The system shows monotonic improvement across generations, achieving Sharpe ratios above 1.5 on a held-out test period.

QuantEvolve also highlights some of the statistical challenges that arise when applying agentic loops to finance. The authors note that ``strategies generated by the framework may be susceptible to data snooping bias'' and that they ``have not formally validated whether generated hypotheses reflect established market theories or provide post-hoc rationalizations for data-mined patterns.'' In our work, in addition to a more in-depth exploration of these optimization settings, we attempt to address the challenge of overfitting systematically. For every experiment, we track out-of-sample performance and explicitly discuss p-hacking and backtest overfitting \citep{bailey2014backtest}. 

\subsection{AI in Quantitative Finance}

The application of machine learning to finance has a long history \citep{lopez2018advances}, but the integration of LLMs into trading systems is nascent. Early work showed that even zero-shot LLM sentiment scores carry predictive power for stock returns \citep{lopezlira2023}, while subsequent efforts have focused on fine-tuned financial LLMs \citep{zhang2023fingpt, zhang2023instructfingpt} and multi-agent trading simulations \citep{li2023tradinggpt, xiao2024tradingagents}. \citet{lee2024finllms} survey the landscape, noting both promise and significant challenges around hallucination, data privacy, and efficiency.

More recently, two developments signal a shift from LLMs-as-components toward LLMs-as-decision-makers. \citet{alphaagents2025} demonstrate a multi-agent architecture for equity portfolio construction at BlackRock, where specialized LLM agents perform fundamental, sentiment, and valuation analysis under explicit risk-tolerance profiles. And the AIA Forecaster from Bridgewater AIA Labs \citep{aiaforecaster}---an agentic system combining adaptive search, supervisor-led reconciliation, and statistical calibration---achieved performance statistically indistinguishable from human superforecasters on the ForecastBench benchmark, providing the first institutional-scale evidence that agentic architectures can match expert-level financial judgment.

\subsection{Two Problem Classes: Forecasting vs.\ Algorithm Optimization}
\label{sec:taxonomy}

We distinguish two fundamentally different problem classes that arise in trading system optimization: Financial Forecasting and Algorithm Optimization. Each has a different reward structure, a different failure mode, and a different relationship to the p-hacking problem.

\subsubsection{Financial Forecasting}
\label{sec:forecasting_problem}

The forecasting problem is: \emph{predict a target $Y$ (future returns, volatility, or other quantities) from available market data}. The agentic loop may construct features from raw data, select among existing indicators, design the model architecture (e.g., the structure of a neural network or the specification of a factor model), iterate on the fitting procedure, or any combination of these. The reward function is a measure of predictive accuracy---typically $R^2$, information coefficient (IC), or a rank correlation against realized outcomes.

This is a noisy optimization problem par excellence. True predictive signals in financial data are small: cross-sectional $R^2$ values of 1--3\% are considered meaningful, and even the best known alpha factors exhibit Sharpe ratios that imply signal-to-noise ratios well below 1. An agentic loop that evolves forecasting models is therefore searching for needles in a haystack of noise, and the risk of ``discovering'' spurious patterns is extreme.

The key concern is that \emph{the loop itself is a multiple testing procedure}. Each candidate forecaster evaluated against historical data constitutes an independent hypothesis test. After $N$ iterations, the best in-sample $R^2$ is biased upward by an amount that grows with $N$---this is precisely the mechanism underlying the ``backtest overfitting'' phenomenon analyzed by \citet{bailey2014backtest}. If the loop is merely generating and testing random model variants, it is no different from running $N$ regressions and picking the best one: a sophisticated form of p-hacking.

However, there is a crucial distinction: \emph{agentic loops do research, not just search}. An LLM proposing a new forecasting model is not sampling uniformly from the space of possible regressions. It brings economic reasoning, knowledge of factor literature, and the ability to synthesize information from multiple sources. 
If the LLM's proposals are systematically better than random---if they are informed by genuine financial intuition---then the search should follow a qualitatively different trajectory from classical overfitting. In the pure overfitting case, the OOS curve should remain flat or deteriorate as the loop keeps selecting better IS candidates; in the agentic case, genuinely useful mutations should produce candidates whose OOS performance improves alongside their IS performance. Testing for this difference in trajectory is a central goal of our experiments.

\subsubsection{Algorithm Optimization: Execution and Portfolio Construction}
\label{sec:algo_problem}

The second class of problems involves optimizing \emph{algorithms}---execution strategies, portfolio constructors, risk models---rather than forecasts. Here the reward function is typically PnL or Sharpe ratio computed via a backtest simulation.

These problems differ from forecasting in several important ways:

\begin{itemize}
    \item \textbf{Larger impact of individual mutations}:
    Mutations to an execution or portfolio algorithm often make large, discrete changes to trading behavior that have an immediate and visible effect on PnL and Sharpe, whereas mutations to a forecasting pipeline often add features that are redundant with existing ones or improve fit only marginally.
    \item \textbf{More direct feedback from the objective}: For fixed market data, both forecasting metrics and execution PnL are deterministic. However, an important difference is that execution and portfolio-construction mutations are judged directly on the ultimate objective---PnL or Sharpe---and their effects are often immediately visible. In forecasting, by contrast, one can still make progress in $R^2$ or IC through new features without seeing immediate progress in PnL, especially when new features are highly correlated with existing ones or when the problem is already near diminishing returns. This makes progress in execution optimization easier to detect and act on.
    \item \textbf{The algorithm is conditional on a financial forecast}: The quality of execution and portfolio construction is always measured \emph{given} a set of trading signals. This factorization means we can hold the forecast fixed and vary the algorithm, isolating algorithmic improvement from signal quality.
    \item \textbf{Different overfitting risk}: The primary risk is not p-hacking in the statistical sense but rather \emph{overfitting to the assumptions of the backtest itself}. If the simulation does not accurately reflect production conditions, optimized results will not transfer. This is particularly acute for position sizing: without realistic market impact modeling \citep{bouchaud2009impact}, the ``optimal'' portfolio size is unrealistically large, and an algorithm tuned against a frictionless backtest will oversize positions systematically. 
    We address this challenge by including a market impact in the PnL calculation (see App. \ref{ssec:market_impact} for details). A secondary risk is \emph{overfitting to a particular market regime}---an execution algorithm optimized on 2024's volatility regime may fail in a 2025 regime shift. We partially address this by keeping an independent test set that extends over 9 months and confirming that performance gains persist across extended periods of time.  
\end{itemize}

As we will show, algorithm optimization is where agentic loops are most naturally suited: the search space is large enough to benefit from automated exploration, the evaluation is relatively crisp, and the LLM's ability to generate structurally novel code (not just parameter variations) provides a significant advantage over traditional hyperparameter optimization.

\subsubsection{Joint Optimization of Forecasting and Algorithm}
\label{sec:joint_problem}
The most ambitious---and most interesting---configuration is \emph{joint evolution} of forecaster, optimizer, and executor simultaneously. In principle, jointly optimizing the entire pipeline should capture interaction effects that component-level optimization misses: a forecasting signal that is ``bad'' in isolation may be excellent when paired with the right execution algorithm, or vice versa.

In practice, joint optimization dramatically expands the search space and makes the reward signal noisier (since changes to any component affect PnL). It also complicates the attribution of improvement: when PnL increases, is it because the forecast improved, the algorithm improved, or some combination? We study whether the benefits of joint optimization outweigh these costs in two complementary settings: jointly evolving the two execution stages while holding the forecaster fixed (Section~\ref{sec:exp_joint}), and jointly evolving the forecaster's feature pipeline together with the full execution logic (Section~\ref{sec:exp_joint_calcset_strategy}).

\section{Brief Review of the MadEvolve Framework}
\label{sec:framework}
\textsc{MadEvolve} is a general-purpose framework for LLM-driven evolutionary code optimization, inspired by FunSearch~\citep{romera2024funsearch} and AlphaEvolve~\citep{alphaevolve2025} (see also the open-source reimplementations ShinkaEvolve~\citep{lange2025shinkaevolve} and OpenEvolve~\citep{OpenEvolve2025}), and described in full detail in~\citet{li2026madevolve}, where it was applied to three problems in computational cosmology. The framework is publicly available at \url{https://madevolve.org}. Here we summarize the components relevant to the present work and note the adaptations made for the trading domain.

\subsection{Evolution Loop}

\begin{figure}[h]
    \centering
    \includegraphics[width=0.7\textwidth]{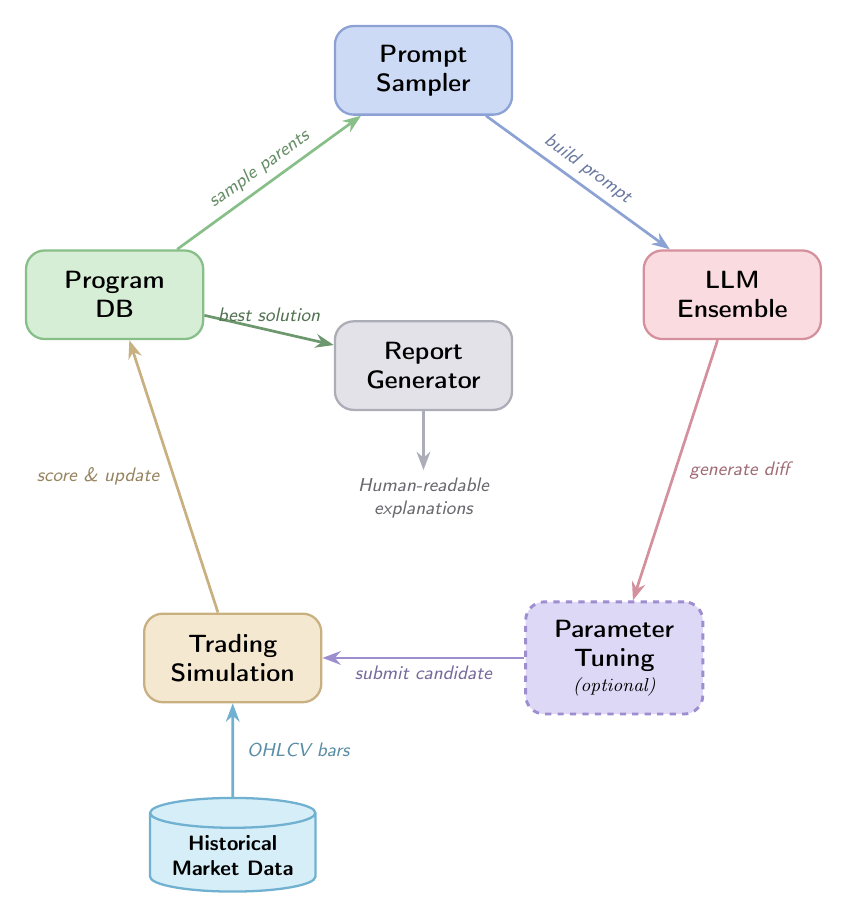}
    \caption{Overview of the \textsc{MadEvolve} evolution loop. The prompt sampler retrieves parent and inspiration programs from the population database, queries the LLM ensemble, evaluates the resulting candidate against the backtester, and updates the population. See~\citet{li2026madevolve} for the full architecture.}
    \label{fig:framework_overview}
\end{figure}

The framework operates as a closed-loop system (Figure~\ref{fig:framework_overview}). Each iteration proceeds through five stages: (i)~a \emph{parent program} is sampled from a structured population database, balancing exploitation of high-fitness solutions against exploration via inter-island migration; (ii)~a set of \emph{inspiration programs}---including the current global best, recent top performers, and structurally diverse neighbors---is retrieved from the database; (iii)~a prompt containing the parent source code, its performance metrics, and the inspirations is assembled and sent to an LLM, which returns either a targeted diff patch or a complete rewrite of the mutable code region; (iv)~the candidate is \emph{evaluated} against the task-specific fitness function (in our case, a minute-bar backtest simulation returning impact-adjusted PnL); and (v)~the scored candidate is inserted into the population database, updating the MAP-Elites grid and island archives.

The user designates which portions of the source code are mutable by enclosing them in \texttt{EVOLVE-BLOCK-START} / \texttt{EVOLVE-BLOCK-END} markers. All code outside these markers---including the simulation harness, data loading, and PnL accounting---remains fixed throughout evolution, ensuring that fitness improvements reflect genuine algorithmic changes rather than evaluation artifacts.

\subsection{Population Management}

Maintaining a diverse population is critical for avoiding premature convergence in large search spaces. This diverse population is a possible advantage over other agentic systems such as Claude Code (without additional structure), as we will discuss further in Sec. \ref{sec:claudecode}. \textsc{MadEvolve} combines three complementary mechanisms from the evolutionary computation literature, largely following ~\citep{alphaevolve2025}.

A \emph{MAP-Elites grid}~\citep{mouret2015mapelites} partitions the behavioral space into cells indexed by code complexity (character length), diversity (mean cosine distance of the program's embedding to a reference set), and performance score. Each cell retains only its best-performing occupant, ensuring that the population covers a range of structurally distinct solutions even as selection pressure drives fitness upward.

An \emph{island model} partitions the population into semi-isolated subpopulations that evolve independently. Periodic migration events transfer a fraction of each island's elites to its neighbor via a ring topology, enabling cross-pollination of successful strategies without homogenizing the population. We use five islands with migration every five generations at a 10\% transfer rate.

A \emph{global elite archive} of fixed capacity preserves the highest-scoring programs regardless of island membership, preventing loss of top solutions during migration or grid replacement.

\subsection{LLM Ensemble and Model Selection}
\label{sec:framework_llm}

The framework supports an ensemble of LLM providers (OpenAI, Anthropic, Google, DeepSeek) through a unified gateway. Rather than committing to a single model, \textsc{MadEvolve} routes mutation queries across the available models, which exhibit complementary strengths: lightweight models (e.g., Gemini Flash) provide throughput for incremental refinement, while more capable models (e.g., Claude Opus) occasionally propose deeper structural innovations. The ensemble supports three generation modes: \emph{differential patches} (${\sim}70\%$ of queries) for localized edits, \emph{full rewrites} (${\sim}30\%$) for larger structural departures, and \emph{synthesis} of ideas from multiple parents, analogous to crossover in genetic algorithms.

\subsection{Adaptations for Trading}

Several aspects of the framework were adapted for the trading optimization setting relative to the cosmological applications in~\citet{li2026madevolve}. In the cosmological applications, fitness is defined by physics-based metrics (e.g., cross-correlation coefficients evaluated on simulation grids), and the framework includes an inner optimization loop that tunes continuous parameters via auto-differentiation before computing fitness. We adapted these properties to finance as follows. 

\paragraph{Fitness function.} 

We use impact-adjusted PnL on the validation period as the fitness signal for runs that evolve trading code, computed through the backtest simulation described in Section~\ref{sec:trading_setup}. The forecasting-only Run~4 (Section~\ref{sec:run4}) is the exception: there we optimize a composite of $R^2$, mean IC, and ICIR (Equation~\ref{eq:run4_score}), since the evolved features are scored by predictive metrics rather than by simulator output. We deliberately optimize impact-adjusted PnL rather than the Sharpe ratio for two reasons. First, Sharpe ratio can be inflated by trading more selectively---picking only the highest-confidence signals or sitting out noisy periods---which would bias the search toward strategies that trade too little. Second, the impact-adjusted PnL already embeds a calibrated trading-cost penalty that internalizes the risk-aversion role Sharpe typically plays, while preserving a direct economic incentive to actually generate returns at scale. The symmetric concern---that the loop could inflate PnL by simply scaling positions up rather than improving the underlying algorithm---is bounded by the super-linear impact term derived in Appendix~\ref{ssec:market_impact}, which makes the fitness function eventually concave in trade size; we further verify in Section~\ref{subsec:summary} (Figure~\ref{fig:pnl_per_volume}) that the evolved strategies beat a pure-sizing counterfactual of the baseline in every run, and that the scale-invariant Sharpe and Calmar ratios both improve out-of-sample for every run, three signatures jointly inconsistent with a pure scale-up of the baseline.

\paragraph{Parameter tuning.} Unlike in cosmological applications, we do not employ a separate inner parameter optimization loop; instead, all tunable quantities---multipliers, thresholds, decay constants---are embedded directly in the evolved code and modified by the LLM alongside the algorithmic logic. This is a deliberate choice rather than an oversight. Re-fitting each candidate's continuous parameters to the validation period at every generation would complicate the trial-count factor discussed below, since the effective number of trials per candidate would no longer be a handful of LLM mutations but a continuous sweep of each parameter's local optimum, feeding directly into the multiple-testing inflation analyzed in Section~\ref{sec:discussion} and biasing the search toward over-parameterized code structures rather than more robust ones. The trading objective is also not particularly gradient-friendly: the strategy is dominated by discrete branching---side decisions, risk-reduction toggles, threshold gates---and the backtest contains discrete fill events whose contribution to PnL is piecewise in the limit price, so a faithful gradient through the rollout would have to cope with non-differentiabilities at every step. A differentiable surrogate of the simulator could in principle re-enable an inner loop under a relaxed objective, and we leave this to future work. We do, however, experiment with Bayesian parameter optimization in Sec. \ref{sec:hyperparam_calibration}.

\paragraph{Domain-specific prompting.} The prompt adapter injects trading-specific context: a description of the alpha signal semantics, the limit order execution model, the PnL decomposition, and constraints on position size and order frequency. This grounding helps the LLM propose modifications that respect the structure of the problem rather than treating the code as a generic optimization target. 

\paragraph{Parameter budget.} To limit the effective complexity of the evolved code, we cap the number of named tunable parameters per candidate program---typically 15--20, set per run depending on the size of the \texttt{EVOLVE-BLOCK}---and require that every tunable quantity be declared as an \texttt{UPPER\_CASE} module-level constant at the top of the \texttt{EVOLVE-BLOCK}, so that the parameter count is unambiguous and easy to inspect. Candidates exceeding the cap are penalised, which encourages the LLM to express new ideas as algorithmic structure rather than as additional free parameters. We adopted this constraint after early pilot runs without it produced strategies that fit the validation period strongly but failed to generalize out-of-sample, in line with the multiple-testing inflation analyzed in Section~\ref{sec:discussion}: every additional free parameter expands the effective search space and inflates the gap between in-sample and out-of-sample performance. The same naming convention is used across all evolutionary runs reported in this paper.

\paragraph{Report generation.} \textsc{MadEvolve} automatically generates human-readable reports analyzing the best evolved strategy relative to the baseline. The report traces the evolutionary lineage, identifies structural innovations, and provides a narrative comparison grounded in trading-specific concepts (adverse selection, fill probability, inventory risk). We found these reports invaluable for rapidly assessing whether a high-scoring candidate reflects a genuine algorithmic insight or an overfitting artifact. Excerpts are shown in App.~\ref{app:run_report_summaries}.

\section{Trading Simulation Setup and Base Strategy}
\label{sec:trading_setup}

We now discuss our trading simulation, as well as the base strategy from which we evolve the forecaster and algorithm. We chose a passive limit order strategy on Bitcoin market data, since crypto trading is relatively accessible and well-contained.

\subsection{Trading Simulation Setup}
\label{sec:tradingsim}

We evaluate all candidate strategies in a unified minute-bar backtesting environment for BTCUSD. The setup is designed to isolate strategy logic while keeping execution assumptions and data splits fixed across all experiments. 

For all runs, we use BTCUSD 1-minute OHLCV bars from \texttt{polygon}\footnote{https://massive.com/docs/rest/crypto/overview}, and split time chronologically into train (2022--2023), validation (2024), and test (2025 through 2025-10-10). The train split is used only for alpha model fitting, while the evolutionary loop optimizes on validation performance and reports out-of-sample generalization on test.

At a high level, the simulator models passive limit-order execution with one decision per minute, computes net and impact-adjusted PnL, and applies a calibrated market-impact penalty to discourage unrealistically aggressive trading. This defines a single, programmatic reward signal for the agentic loop.

Operationally, each minute the strategy proposes a signed trade size and a limit price, the simulator checks whether that resting order would have filled from the realized candle range, updates inventory, and then replaces the prior order with a new one. This one-order-at-a-time lifecycle keeps execution logic interpretable while still capturing the key interaction between signal quality, position management, execution style, and trading frictions. Throughout the paper, candidate programs are ranked by impact-adjusted PnL on the validation split and then evaluated unchanged on the held-out test split. Full mathematical and implementation details (fill logic, order lifecycle, PnL decomposition, and impact parameterization) are provided in Appendix~\ref{app:simulation_details}.

We emphasize that the described backtest setup is still far from being realistic. Minute-bar data from \texttt{polygon} is not exchange-specific, but rather aggregated across different exchanges. Therefore, we do not expect that our quantitative results will hold "out-of-the box" on any real exchange, since we don't model exchange-specific operational details. Nevertheless, this work is focused on a proof-of-concept and an executional realism is left for future exploration.

\subsection{Base Strategy}
\label{sec:basestratexplain}
In this section, we outline the core logic elements of the strategy and forecaster, and refer to the Appendix \ref{app:basestrat} for more implementational details and a skeleton code. The baseline strategy combines a fixed alpha forecaster with a two-stage decision pipeline. The forecaster is a ridge regression over three demeaned exponential moving averages of one-step returns at halflives of 1, 5, and 10 minutes, fit on the 2022--2023 training split. The default strategy takes a short-term prediction $\alpha$, converts it into a desired BTC position, and then places a single limit order to move the current position toward that target. The strategy calculates two elements: the desired target position and the limit price of the corresponding order. The target position is calculated as follows. Given the predicted signal strength, $\alpha$, and it's standard deviation, $\sigma_\alpha$, the strategy forms cost-adjusted long and short target positions,
\[
q_{\mathrm{long}} = \frac{A}{\sigma_\alpha}\,(\alpha - f_{\mathrm{exp}}),
\qquad
q_{\mathrm{short}} = \frac{A}{\sigma_\alpha}\,(\alpha + f_{\mathrm{exp}}).
\]
where $A$ is a position sizing factor and $f_\mathrm{exp}$ is a fee threshold. Both quantities are parameters of the strategy. Let $q_\mathrm{usd}$ be a current position in USD. If $q_{\mathrm{long}} > q_{\mathrm{usd}}$, the strategy moves toward the long target; if $q_{\mathrm{short}} < q_{\mathrm{usd}}$, it moves toward the short target; otherwise it keeps the current position. In other words,
\[
q_{\mathrm{target}} =
\begin{cases}
q_{\mathrm{long}}, & q_{\mathrm{long}} > q_{\mathrm{usd}},\\[0.3em]
q_{\mathrm{short}}, & q_{\mathrm{short}} < q_{\mathrm{usd}},\\[0.3em]
q_{\mathrm{usd}}, & \text{otherwise.}
\end{cases}
\]
This raw target is subsequently adjusted by several heuristic correction factors, including stale-signal and inventory-based corrections; details are given in the skeleton code in App. \ref{app:skeleton_code}. The trade quantity is then clipped to the maximum allowed order size. If its USD value is below the minimum trade threshold, no order is placed. If an order is placed, the strategy uses exponential price scaling around the mid price. Writing $d$ for the chosen order depth (i.e. how far in the book is the order placed), the limit price is
\[
p_{\mathrm{limit}} = m\, e^{-s d},
\]
where $s=+1$ for a buy order and $s=-1$ for a sell order.

\section{Results}
\label{sec:results}

We first conduct three runs of MadEvolve corresponding to different scopes of the algorithm evolution (execution and portfolio construction), using the fixed baseline forecaster described in Section~\ref{sec:trading_setup}:

\begin{enumerate}
    \item \textbf{Run~1}: Only the target computation (\texttt{set\_target}) is evolved; order placement is inherited from the base class.
    \item \textbf{Run~2}: Only the order placement (\texttt{set\_limit\_order}) is evolved; the target computation is inherited.
    \item \textbf{Run~3}: The full strategy pipeline (\texttt{set\_passive\_order\_data}), encompassing both stages, is evolved jointly.
\end{enumerate}

In addition, we evolved the financial forecast (alpha-predictor), first alone and then jointly with the execution logic:
\begin{enumerate}
    \setcounter{enumi}{3}
    \item \textbf{Run~4}: Only the feature set (\texttt{default\_calcset}) feeding the fixed ridge alpha predictor is evolved; the strategy and execution logic are inherited from the baseline and the predictor is refit from scratch on every candidate.
    \item \textbf{Run~5}: Both the feature set (\texttt{default\_calcset}) and the full execution pipeline (\texttt{set\_passive\_order\_data}) are evolved jointly; the ridge predictor is refit on evolved features at every candidate, and the evolved strategy consumes the resulting alpha. This is the full \enquote{cross-product} of Runs~3 and~4.
\end{enumerate}

The five runs are thus roughly in order of difficulty and noisiness of the underlying problem. Skeleton code for the four evolvable functions (\texttt{set\_target}, \texttt{set\_limit\_order}, \texttt{set\_passive\_order\_data}, and \texttt{default\_calcset}) is provided in Appendix~\ref{app:skeleton_code}.

\subsection{Summary of Strategy and Joint Feature+Strategy Results}
\label{subsec:summary}

Table~\ref{tab:summary} presents the headline results across the four evolution runs that ultimately optimize impact-adjusted PnL through the backtester (Runs~1, 2, 3, and~5). Run~4 is omitted from this summary because its fitness signal is a forecast-quality composite rather than PnL; its results are reported separately in Table~\ref{tab:run4_metrics}. All four runs in Table~\ref{tab:summary} share the same BTCUSD minute-bar data, chronological splits, and PnL accounting; they differ only in the scope of the code block exposed to evolution and, for Run~5, in the underlying feature set feeding the ridge predictor. Although the four runs use slightly different code templates---Runs~1 and~2 leave one component at its baseline implementation, Run~3 evolves a combined single-function variant, and Run~5 pairs the baseline three-feature ridge predictor with that same combined template---the templates are initialised to be functionally equivalent, so all four configurations reduce to the same numerical baseline (4.81 validation Sharpe, \$83K validation impact-adjusted PnL) and Table~\ref{tab:summary} reports a single shared baseline row.

\begin{table}[h]
\centering
\scriptsize
\setlength{\fboxsep}{8pt}
\fcolorbox{mainresultborder}{mainresultbg}{%
\begin{minipage}{0.98\linewidth}
\centering
\setlength{\tabcolsep}{3pt}
\renewcommand{\arraystretch}{1.05}
\begin{tabular}{lcccccc}
\toprule
 & \multicolumn{3}{c}{\textbf{Validation (2024)}} & \multicolumn{3}{c}{\textbf{Test (2025, OOS)}} \\
\cmidrule(lr){2-4} \cmidrule(lr){5-7}
\textbf{Run} & Sharpe & PnL (\$K) & Vol.\ (\$M) & Sharpe & PnL (\$K) & Win \\
\midrule
Baseline (shared) & 4.81 & 83 & 502 & 3.82 & 47 & 60.1\% \\
\midrule
1: Target only & 4.83 (1.00x) & 533 (6.42x) & 3{,}336 (6.65x) & 4.45 (1.16x) & 271 (5.77x) & 55.1\% (0.92x) \\
2: Order only & 6.49 (1.35x) & 2{,}238 (27.0x) & 10{,}289 (20.5x) & 5.12 (1.34x) & 1{,}205 (25.6x) & 61.1\% (1.02x) \\
3: Joint strategy & 6.51 (1.35x) & 973 (11.8x) & 4{,}449 (8.86x) & 5.11 (1.34x) & 473 (10.1x) & 53.0\% (0.88x) \\
5: Joint features+strategy & 8.85 (1.84x) & 1{,}855 (22.3x) & 7{,}318 (14.6x) & 5.65 (1.48x) & 724 (15.4x) & 49.8\% (0.83x) \\
\bottomrule
\end{tabular}
\end{minipage}%
}
\caption{Summary of evolution results across the four PnL-optimizing runs. PnL figures are impact-adjusted. Values in parentheses on evolved rows report the ratio relative to the shared baseline. Although the four runs use slightly different code templates---Runs~1 and~2 expose only one component (target or order placement) while keeping the other at its baseline implementation, Run~3 evolves the combined single-function variant, and Run~5 pairs the baseline three-feature ridge predictor with that same combined template---the templates are initialised to be functionally equivalent, so all four reduce to the same numerical baseline performance and we report a single shared row.}
\label{tab:summary}
\end{table}

\begin{figure}[!htbp]
    \centering
    \includegraphics[width=0.95\textwidth]{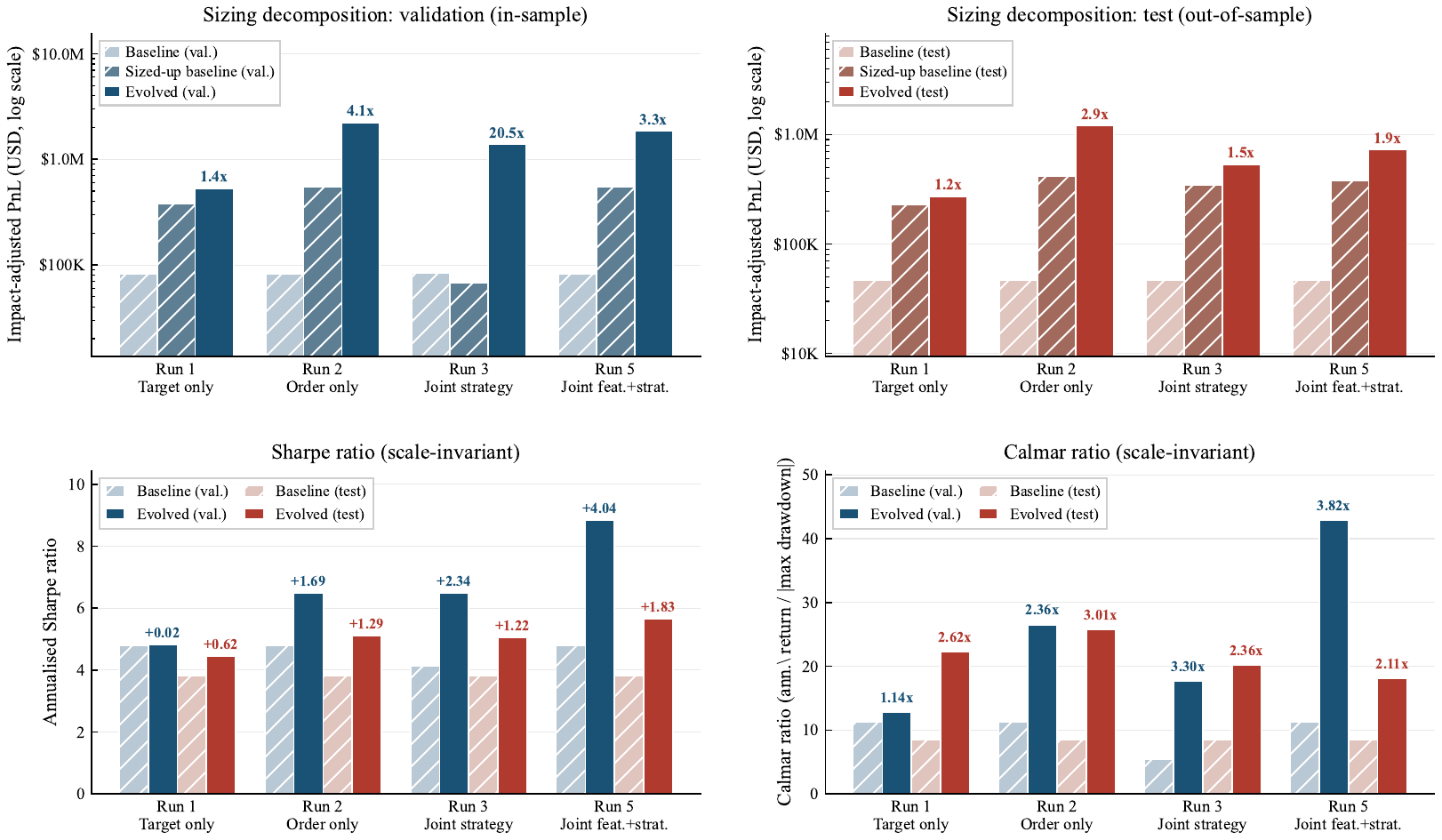}
    \caption{Three complementary tests of the ``is the gain just sizing?'' question. \emph{Top row:} sizing decomposition. The three bars per cluster are baseline impact-adjusted PnL (light grey), the counterfactual PnL the baseline would produce if scaled uniformly to the evolved volume using the impact model of Appendix~\ref{ssec:market_impact} (mid-grey, hatched), and the actual evolved PnL (solid); the annotation above each evolved bar is the ratio of actual to pure-sizing counterfactual, with values above $1.0\times$ giving direct evidence of algorithmic improvement. The $y$-axis is logarithmic. \emph{Bottom row:} Sharpe ratio (left) and Calmar ratio (right), both of which are scale-invariant under a linear rescaling of trade sizes, so any improvement is by construction not attributable to sizing.}
    \label{fig:pnl_per_volume}
\end{figure}

Several patterns emerge from these results. First, all four runs produce substantial improvements in both in-sample and out-of-sample performance, with out-of-sample Sharpe ratios increasing by 0.6 to 1.8 points over their respective baselines. Second, the gains are accompanied by large increases in trading volume; we examine below (Figure~\ref{fig:pnl_per_volume}) how much of this is genuine algorithmic improvement versus simple up-sizing. Third, evolving the order placement alone (Run~2) yields the largest absolute PnL, suggesting that execution optimization contributes disproportionately to net PnL at the frictions assumed by our backtester. Fourth, and perhaps most surprisingly, the joint strategy run (Run~3) does not dominate the component-wise runs on all metrics despite having access to a strictly larger search space---a pattern we return to in Section~\ref{sec:discussion}. Fifth, the full joint feature+strategy evolution (Run~5) achieves the highest validation and test Sharpe ratios in the entire set (8.85 and 5.65, respectively), demonstrating that letting evolution co-design the feature pipeline and the execution logic can capture interactions that either axis alone cannot. However, Run~5 also shows the largest validation-to-test PnL retention gap (roughly 39\%, against 49--54\% for Runs~1--3) and the largest win-rate drop (68.6\% $\to$ 49.8\%), consistent with the enlarged search space amplifying both genuine discoveries and overfitting pressure.

\paragraph{Sizing vs.\ algorithmic improvement.}

Because impact-adjusted PnL is monotone in trade size up to the point at which impact dominates, a natural concern is that the agent might be inflating PnL primarily by sizing positions up rather than by improving the underlying algorithm. We attack the question from three complementary angles, all of which point in the same direction. The results are summarised in Figure~\ref{fig:pnl_per_volume}.

\emph{(i) Sizing-only baseline.} Using the impact model of Appendix~\ref{ssec:market_impact}, we can ask what the baseline strategy would have earned if its trades had been scaled up by $k = V_e / V_b$, the volume ratio between the evolved and baseline runs. Frictionless PnL grows linearly in $k$ and impact cost as $k^{1.5}$ (per-trade displacement scales as $|q|^{0.5}$), giving $\mathrm{PnL}_{\mathrm{sized}}(k) = k\,F_b - k^{1.5}\,I_b$, with $F_b$ and $I_b$ the baseline's frictionless PnL and impact cost. The ratio $\mathrm{PnL}_{\mathrm{evolved}} / \mathrm{PnL}_{\mathrm{sized}}$ then isolates the part of the gain that pure rescaling cannot account for: it equals $1$ if the agent only sized up, and is larger otherwise. We find this ratio above $1$ on both splits in every run (Figure~\ref{fig:pnl_per_volume}), in the $1.2$--$2.9\times$ range on test and the $1.4$--$4.1\times$ range on validation. A simpler per-dollar-traded efficiency would not separate these effects, since evolved strategies tend to place fewer, larger trades whose per-dollar impact lies at a different point on the $k^{1.5}$ curve.

\emph{(ii) Sharpe ratio.} A linear scaling of every trade by a constant factor leaves Sharpe unchanged, and adding the super-linear impact term can only push sized-up Sharpe \emph{down}; Sharpe is therefore an entirely scale-invariant test of strategy quality. Out-of-sample Sharpe improves in all four runs, by $+0.62$ for Run~1, $+1.29$ for Run~2, $+1.29$ for Run~3, and $+1.83$ for Run~5 (Figure~\ref{fig:pnl_per_volume}, lower-left). The in-sample Sharpe gain is essentially zero for Run~1 ($+0.02$) but materially positive for the other three runs ($+1.69$ for Run~2, $+1.70$ for Run~3, and $+4.04$ for Run~5); read together with the counterfactual-sizing ratio, this says that Run~1's in-sample improvement is mostly sizing while its out-of-sample improvement is genuine, and that Runs~2, 3, and 5 are doing real algorithmic work in both regimes.

\emph{(iii) Calmar ratio.} Calmar (annualised return divided by the absolute maximum drawdown) is also scale-invariant under linear scaling. It rises in every run on both splits, with out-of-sample multipliers of $2.6\times$, $3.0\times$, $2.5\times$, and $2.1\times$ for Runs~1, 2, 3, and~5 respectively (Figure~\ref{fig:pnl_per_volume}, lower-right). Because Calmar penalises tail risk explicitly, the across-the-board improvement also rules out the related concern that the evolved strategies are simply trading drawdown depth for headline PnL.

A fourth, independent check comes from the qualitative changes in the evolved code itself (described in Sections~\ref{sec:exp_joint}, \ref{sec:exp_joint_calcset_strategy}, and Appendix~\ref{app:run_report_summaries}): the evolved strategies introduce structural innovations such as signal-conviction-dependent sizing, inventory-aware quoting depth, regime-conditional risk reduction, and superlinear conviction boosts. These changes cannot be reproduced by any single sizing constant.

\subsection{Run 1: Evolving the Target Position Based on Alpha}

The first experiment restricts evolution to the target computation, holding the order placement logic fixed at its baseline implementation. The evolution ran for approximately 14.6 hours, evaluating 990 candidate programs across 334 generations, with the best variant discovered after roughly 962 programs (generation 324).

Qualitatively, the best evolved target logic replaces the baseline's stateless thresholding with a more stateful signal-processing rule that reacts to momentum, inventory, and trade urgency. A detailed human-edited summary of the AI-generated run report is provided in Appendix~\ref{app:run_report_summaries}, under Run~1.

\paragraph{Performance.}
On the validation set, the evolved strategy increases impact-adjusted PnL from \$83K to \$533K while maintaining a Sharpe ratio of 4.83. The strategy achieves this primarily through a $6.6\times$ increase in trading volume, driven by a sizing scale multiplier of 2.46. On the held-out test set, the Sharpe ratio improves from 3.82 to 4.45, and impact-adjusted PnL increases from \$47K to \$271K. The win rate declines from 60.1\% to 55.1\% out-of-sample, reflecting a shift toward capturing larger moves at the cost of precision on marginal trades. Maximum drawdown increases from \$8.2K to \$22.5K in absolute terms but remains proportionate to the much larger PnL.

\begin{figure}[h]
    \centering
    \includegraphics[width=\textwidth]{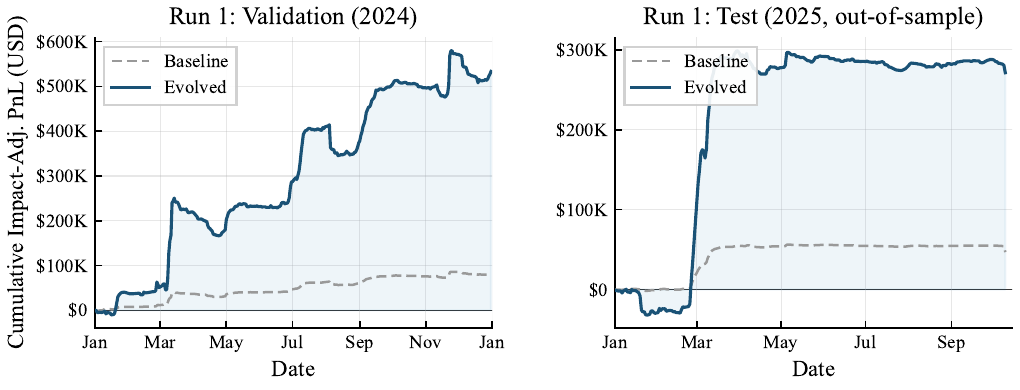}
    \caption{Cumulative impact-adjusted PnL for the baseline and best evolved strategy in Run~1 (target position evolution). Left: validation set (2024). Right: test set (2025).}
    \label{fig:run1_pnl}
\end{figure}

\begin{figure}[h]
    \centering
    \includegraphics[width=0.6\textwidth]{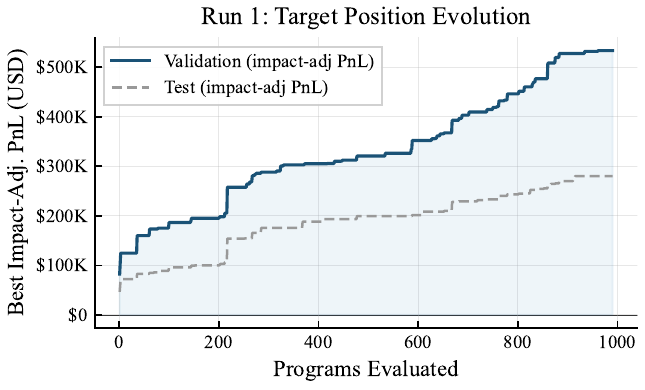}
    \caption{Evolution progress for Run~1. Highest impact-adjusted PnL achieved by any candidate program in the first $n$ programs evaluated, on the validation (in-sample) and test (out-of-sample) splits.}
    \label{fig:run1_progress}
\end{figure}

\subsection{Run 2: Evolving the Order Strategy Based on Target Position}

The second experiment evolves only the order placement logic while keeping the target computation at its baseline. This run executed for approximately 19.1 hours over 334 generations, evaluating 997 candidate programs, with the best variant appearing after roughly 972 programs (generation 325).

Qualitatively, the best evolved execution policy moves from fixed quoting heuristics toward a more adaptive limit-order policy that conditions on fill probability, adverse-selection risk, and signal alignment. A detailed human-edited summary of the AI-generated run report is provided in Appendix~\ref{app:run_report_summaries}, under Run~2.

\paragraph{Performance.}
This run produces the largest absolute gains of the three algorithm experiments. Validation impact-adjusted PnL rises from \$83K to \$2.24M, with the Sharpe ratio increasing from 4.81 to 6.49. On the test set, impact-adjusted PnL reaches \$1.20M and the Sharpe ratio improves to 5.12. These gains are accompanied by a $20\times$ increase in traded volume, from \$502M to \$10.3B. Despite this dramatic scaling, the maximum drawdown remains moderate relative to PnL: \$146K in-sample and \$99K out-of-sample. The win rate holds relatively stable at 63.7\% in-sample and 61.1\% out-of-sample. The strong out-of-sample generalization suggests that the microstructure features and optimization-based execution capture genuine market structure rather than noise.

\begin{figure}[h]
    \centering
    \includegraphics[width=\textwidth]{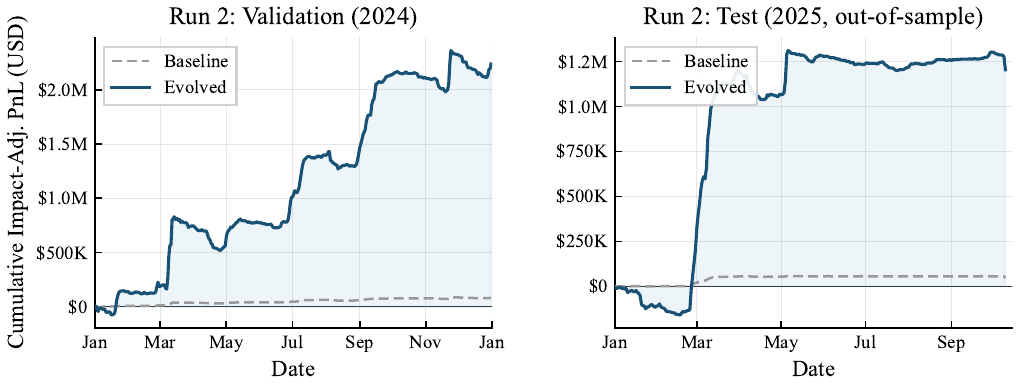}
    \caption{Cumulative impact-adjusted PnL for the baseline and best evolved strategy in Run~2 (order placement evolution). Left: validation set (2024). Right: test set (2025).}
    \label{fig:run2_pnl}
\end{figure}

\begin{figure}[h]
    \centering
    \includegraphics[width=0.6\textwidth]{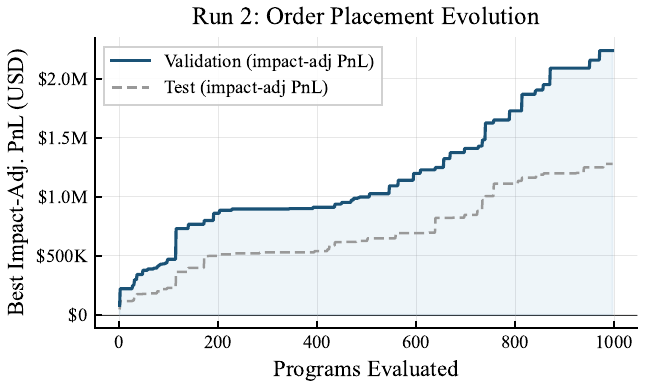}
    \caption{Evolution progress for Run~2. Highest impact-adjusted PnL achieved by any candidate program in the first $n$ programs evaluated, on the validation (in-sample) and test (out-of-sample) splits.}
    \label{fig:run2_progress}
\end{figure}

\subsection{Run 3: Evolving the Target Position and Order Strategy Jointly}
\label{sec:exp_joint}

The third experiment exposes the full decision pipeline to evolution, allowing the optimizer to jointly modify both target computation and order placement within a single code block. The run evaluated 1{,}059 programs across 178 generations, with the best variant found after roughly 984 programs (generation 165).

Qualitatively, the best joint strategy couples signal processing, sizing, and pricing much more tightly than the baseline, with stronger inventory-awareness and regime-dependent execution. A detailed human-edited summary of the AI-generated run report is provided in Appendix~\ref{app:run_report_summaries}, under Run~3.

\paragraph{Performance.}
The joint evolution achieves validation impact-adjusted PnL of \$973K and a Sharpe ratio of 6.51, with an $8.9\times$ increase in volume from \$502M to \$4.45B. On the test set, impact-adjusted PnL reaches \$473K with a Sharpe of 5.11. While these figures represent substantial improvements over the baseline, the joint run underperforms Run~2 in absolute PnL despite having access to a strictly larger search space. The win rate drops more noticeably, from 60.1\% to 53.0\% out-of-sample, suggesting a shift from high-frequency scalping toward a more selective profile that accepts a lower hit rate in exchange for larger gains per trade. The number of evolvable constants in the strategy roughly quadruples from the baseline to the evolved variant.

\begin{figure}[h]
    \centering
    \includegraphics[width=0.6\textwidth]{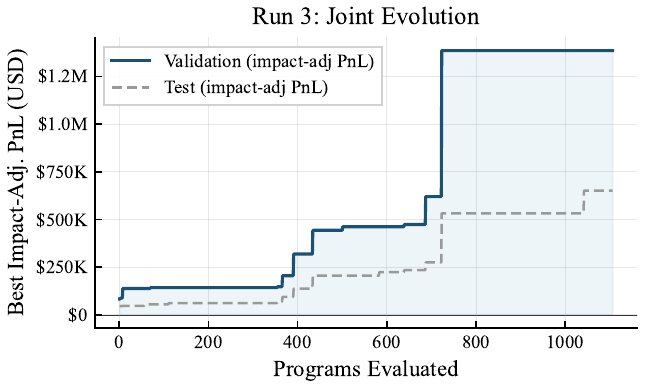}
    \caption{Evolution progress for Run~3. Highest impact-adjusted PnL achieved by any candidate program in the first $n$ programs evaluated, on the validation (in-sample) and test (out-of-sample) splits.}
    \label{fig:run3_progress}
\end{figure}

The evolution progress for this run is notably less monotonic than the component-wise experiments. While Figure~\ref{fig:run3_progress} plots only the cumulative-best score---which is non-decreasing by construction---the per-generation best score oscillates sharply across the run, with the best impact-adjusted PnL within a generation occasionally collapsing to near zero before later programs recover. This pattern suggests that joint optimization of the full pipeline presents a more rugged fitness landscape, where structural changes to one component can temporarily destabilize another before co-adaptation restores coherent behavior.

\begin{figure}[h]
    \centering
    \includegraphics[width=\textwidth]{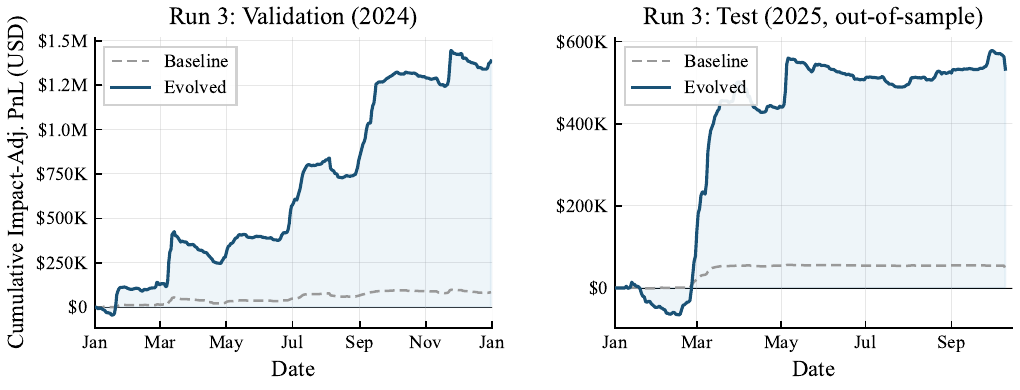}
    \caption{Cumulative impact-adjusted PnL for the baseline and best evolved strategy in Run~3 (joint evolution). Left: validation set (2024). Right: test set (2025).}
    \label{fig:run3_pnl}
\end{figure}

\subsection{Run 4: Evolving the Features for Alpha Prediction}
\label{sec:run4}

In this experiment, we explored the utility of the \textsc{MadEvolve} framework for building a better forecasting model. In all previous runs, the alpha signal driving the strategy is generated by a fixed ridge regression on three EMA-based momentum features---demeaned exponential moving averages of one-step returns at halflives of 1, 5, and 10 minutes. The idea we explore here is to find a more informative feature set for that predictor. Given the vast knowledge of financial literature and demonstrated reasoning skills of frontier LLMs, we expected agentic search to perform reasonably well on this task. The run evaluated 775 candidate programs across 125 generations over approximately 8.5 hours of wall time, with the best variant discovered at generation 100 and not displaced over the subsequent 25 generations.

\paragraph{Setup and scoring function.}
The mutable region is restricted to the body of \texttt{default\_calcset()} (defined in App. \ref{ssec:baseline_alpha}), the function that maps a raw OHLCV dataset to the feature matrix consumed by the alpha predictor. The downstream model---a ridge regression with $\alpha = 0.5$ predicting cumulative log returns at horizons of 1, 10, 100, and 1000 minutes---is held fixed and refit from scratch on every candidate evaluation, using the same train (2022--2023), validation (2024), and test (2025) splits as the strategy runs of Section~\ref{sec:trading_setup}. A single evaluation requires only a ridge fit and predict, with no backtest, and completes in roughly five seconds, which permits a substantially higher generation throughput than the strategy runs at comparable wall time. Impact-adjusted PnL is not a viable fitness signal for feature evolution in this setting: the alpha is consumed by the same fixed execution logic across all candidates, so PnL is dominated by interactions with the existing position-sizing constants rather than by the underlying predictive quality of the features. We instead optimize a weighted combination of three prediction-quality metrics evaluated on the validation split,
\begin{equation}
\label{eq:run4_score}
\text{score} = 0.4 \, R^2_{\text{val}} \;+\; 0.3 \, \text{IC}_{\text{val}} \;+\; 0.3 \cdot \frac{\text{ICIR}_{\text{val}}}{5},
\end{equation}
where $R^2_{\text{val}}$ is the no-intercept out-of-sample $R^2$ at the primary 10-minute horizon, $\text{IC}_{\text{val}}$ is the mean of the daily Spearman rank correlations between predictions and realized returns, and $\text{ICIR}_{\text{val}} = \text{IC}_{\text{val}}/\sigma(\text{IC}_{\text{daily}})$ measures the day-to-day stability of that correlation. Each component is clamped to a bounded range ($R^2$ and IC to $[-1,1]$, ICIR to $[-5,5]$ before its $1/5$ scaling) so that no single metric can dominate the score in pathological cases.

Qualitatively, the evolved feature library broadens the baseline momentum basis into a richer mix of multi-scale momentum, mean-reversion, and volume-sensitive predictors, with explicit clipping for stability. A detailed human-edited summary of the AI-generated run report is provided in Appendix~\ref{app:run_report_summaries}, under Run~4.

\paragraph{Performance.}
Table~\ref{tab:run4_metrics} reports the headline metrics for the baseline and the best evolved feature set on both the validation split (used by the fitness function) and the held-out test split. The combined score improves from $0.085$ to $0.128$, a $51\%$ relative gain, with all three constituent metrics improving in tandem. The mean daily IC rises from $0.074$ to $0.110$ on validation, with the corresponding ICIR improving from $1.03$ to $1.56$. The pattern is preserved on the held-out test split: mean IC rises from $0.059$ to $0.099$ and ICIR from $0.99$ to $1.35$. Validation $R^2$ at the 10-minute horizon roughly doubles from $0.0021$ to $0.0043$, and the analogous test-set $R^2$ also doubles from $0.0017$ to $0.0034$. The in-sample training $R^2$ ($\sim 0.0016$) remains below the validation value, ruling out a naive overfitting failure mode at the model-fit stage; the relevant overfitting question is instead whether the evolutionary loop itself selected features that generalize, which is taken up in Section~\ref{sec:discussion}. The $R^2$ values achieved are rather high. We attribute this to the fact that the \texttt{polygon} dataset is aggregated over multiple exchanges and contains artificial structures that are not present in the any particular exchange data. The evolved set contains 77 columns (versus 3 in the baseline), spanning roughly twenty families: multi-horizon log returns, EMA momentum, realized-volatility ratios, normalized volume, volume-pressure and order-flow signals, candle-shape and wick statistics, Donchian / Bollinger / RSI / MACD variants, VWAP-deviation and mean-reversion terms, choppiness and efficiency ratios, time-of-day periodics, and a few one-off interactions. Absolute pairwise correlations on the 2024 validation split are heavy-tailed (median $0.06$, mean $0.19$), with $16\%$ above $|\rho|=0.5$ and $1.3\%$ above $|\rho|=0.9$. The tail is concentrated within families---average within-family $|\rho|$ reaches $0.91$ for MACD and $0.89$ for realized-volatility---while cross-family values stay below $0.2$. Ridge regularization absorbs this redundancy, which likely explains why the search kept the overlapping horizons rather than pruning them.

\begin{table}[h]
\centering
\small
\setlength{\tabcolsep}{5pt}
\renewcommand{\arraystretch}{1.05}
\begin{tabular}{lcc}
\toprule
\textbf{Metric} & \textbf{Baseline} & \textbf{Evolved} \\
\midrule
Combined score (Eq.~\ref{eq:run4_score}) & 0.0848 & 0.1281 (1.51x) \\
Number of features & 3 & 77 (25.7x) \\
\midrule
\multicolumn{3}{l}{\textit{Validation (2024, in-sample for evolution)}} \\
$R^2$ at 10-min horizon & 0.0021 & 0.0043 (2.05x) \\
Mean daily IC & 0.0736 & 0.1100 (1.49x) \\
ICIR & 1.03 & 1.56 (1.51x) \\
Frictionless PnL proxy (\$) & 214.6 & 498.9 (2.32x) \\
\midrule
\multicolumn{3}{l}{\textit{Test (2025, OOS)}} \\
$R^2$ at 10-min horizon & 0.0017 & 0.0034 (2.00x) \\
Mean daily IC & 0.0592 & 0.0989 (1.67x) \\
ICIR & 0.99 & 1.35 (1.36x) \\
\bottomrule
\end{tabular}
\caption{Run~4 prediction-quality metrics for the baseline three-feature set and the best evolved 77-feature set. Values in parentheses in the evolved column report the ratio relative to baseline. The combined score is the fitness signal optimized by evolution and is computed only on the validation split; test-set metrics are recorded for OOS verification but never seen by the loop. IC denotes the mean daily Spearman rank correlation between predictions and realized returns, and ICIR is its information ratio across days.}
\label{tab:run4_metrics}
\end{table}

\paragraph{Caveats.}
Two caveats are worth noting before reading these numbers as evidence of ``forecast improvement'' in the trading sense. First, $R^2$ values in the $0.003$--$0.005$ range remain very small in absolute terms, which is the default condition for minute-bar return prediction and not a defect of the particular search; the meaningful improvement is the ratio relative to the baseline at the same noise floor. Second, the present run does not put the evolved features back into the trading simulator: a higher mean IC and ICIR do not automatically translate into higher PnL once execution costs and position-sizing dynamics are taken into account. Two natural follow-up experiments address this gap from different angles. The first, discussed in Section~\ref{sec:hyperparam_calibration}, splices the evolved \texttt{default\_calcset} into the baseline strategy and recalibrates the eight execution hyperparameters with Bayesian optimization. The second, which we report in the next subsection, lets evolution co-design the feature set and the full execution logic in a single run, so that features and strategy can adapt to each other rather than being recombined after the fact.

\subsection{Run 5: Joint Evolution of Features and Strategy}
\label{sec:exp_joint_calcset_strategy}

This experiment exposes both \texttt{default\_calcset} and \texttt{set\_passive\_order\_data} to evolution simultaneously, realising the full joint optimisation of forecasting and execution motivated in Section~\ref{sec:joint_problem}. On every candidate evaluation the ridge alpha model is refit from scratch on the evolved feature matrix using the 2022--2023 training split, the resulting alpha signal is fed into the evolved strategy, and the backtester scores the end-to-end pipeline on the 2024 validation period by impact-adjusted PnL. The mutable region spans the entire \texttt{EVOLVE-BLOCK} and contains both functions, so the LLM is free to modify one component per mutation or both at once. The run evaluated 743 candidates across 334 generations over approximately 16 hours of wall time, with the best program found at generation~330 and not displaced over the subsequent four generations.

Qualitatively, the best joint pipeline combines the richer feature construction of Run~4 with an execution policy that inherits the stronger conviction- and inventory-aware logic seen in Runs~1--3. A detailed human-edited summary of the AI-generated run report is provided in Appendix~\ref{app:run_report_summaries}, under Run~5.

\paragraph{Performance.}
Table~\ref{tab:run5_metrics} reports the headline metrics for the baseline and the best evolved pipeline. On the 2024 validation split the impact-adjusted PnL rises from \$83K to \$1.855M, with the Sharpe ratio improving from 4.81 to 8.85---the highest validation Sharpe recorded in any of our evolution runs. On the held-out 2025 test split, the Sharpe ratio improves from 3.82 to 5.65 (again the highest test Sharpe of any run), and impact-adjusted PnL rises from \$47K to \$724K, a $15.5\times$ improvement. Test-set traded volume reaches \$3.6B, roughly half the validation volume (consistent with the shorter 283-day test period), and the test impact-cost rate holds at 0.71~bps, essentially identical to validation.

\begin{table}[h]
\centering
\small
\setlength{\tabcolsep}{5pt}
\renewcommand{\arraystretch}{1.05}
\begin{tabular}{lcc}
\toprule
\textbf{Metric} & \textbf{Baseline} & \textbf{Evolved} \\
\midrule
\multicolumn{3}{l}{\textit{Validation (2024, in-sample for evolution)}} \\
Impact-adjusted PnL (\$) & 82{,}615 & 1{,}854{,}597 (22.4x) \\
Sharpe & 4.81 & 8.85 (1.84x) \\
Max drawdown (\$) & -8{,}739 & -55{,}343 (6.33x) \\
Win rate & 65.3\% & 68.6\% (1.05x) \\
Traded volume (\$M) & 502 & 7{,}318 (14.6x) \\
Impact cost (bps of volume) & 0.32 & 0.72 (2.25x) \\
Number of features & 3 & 79 (26.3x) \\
\midrule
\multicolumn{3}{l}{\textit{Test (2025, OOS)}} \\
Impact-adjusted PnL (\$) & 46{,}791 & 724{,}217 (15.5x) \\
Sharpe & 3.82 & 5.65 (1.48x) \\
Max drawdown (\$) & -8{,}200 & -70{,}091 (8.55x) \\
Win rate & 60.1\% & 49.8\% (0.83x) \\
Traded volume (\$M) & 240 & 3{,}598 (15.0x) \\
Impact cost (bps of volume) & 0.31 & 0.71 (2.29x) \\
\bottomrule
\end{tabular}
\caption{Run~5 performance metrics for the baseline and best evolved joint feature+strategy pipeline. Values in parentheses in the evolved column report the ratio relative to baseline. The baseline is the combined-pipeline strategy fed by the default three-feature ridge predictor. ``Number of features'' counts the columns produced by \texttt{default\_calcset} before the ridge fit.}
\label{tab:run5_metrics}
\end{table}

\paragraph{Interpreting the gains.}
Two observations are worth emphasising. First, Run~5 achieves the best out-of-sample Sharpe of any of our experiments (5.65, versus 4.45, 5.12, and 5.11 for Runs~1--3), despite operating on a strictly larger search space than any individual component run. This cuts against the naive reading of joint optimisation as always \emph{more} prone to overfitting: the Sharpe---which is explicitly scale-invariant---improves, not degrades. Second, the absolute test PnL (\$724K) is smaller than that of the order-placement-only Run~2 (\$1.20M), not because Run~5 is a worse pipeline per unit risk, but because it trades roughly $3\times$ less volume than Run~2 and ends up at a substantially smaller book. Read together, the two observations suggest that the evolved features improve the \emph{quality} of the signal enough that the strategy can afford to be less aggressive about pure volume throughput and still come out ahead on risk-adjusted terms.

\paragraph{Overfitting signature.}
The overfitting signature is nevertheless more pronounced than in the component-wise runs. The validation-to-test impact-adjusted-PnL retention is roughly 39\%, compared with about 51\% for Run~1 and 54\% for Run~2, and the win rate drops from 68.6\% in-sample to 49.8\% out-of-sample---an 18~percentage-point gap, also the largest in the paper. The test-set maximum drawdown, while larger in absolute terms than in the component runs, remains on the order of 10\% of test PnL, so the test-set loss is spread across the period rather than concentrated in a few large drawdowns. The multiple-testing analysis in Section~\ref{sec:discussion} applies here as well, with Run~5's effective trial count of the same order of magnitude as Runs~1--3.

\begin{figure}[h]
    \centering
    \includegraphics[width=0.6\textwidth]{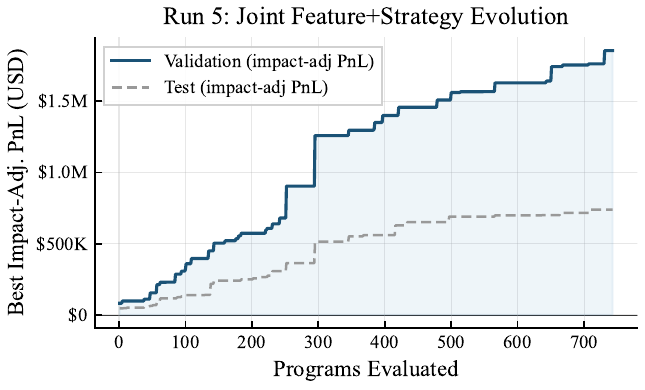}
    \caption{Evolution progress for Run~5. Highest impact-adjusted PnL achieved by any candidate program in the first $n$ programs evaluated, on the validation (in-sample) and test (out-of-sample) splits.}
    \label{fig:run5_progress}
\end{figure}

\paragraph{Evolution dynamics.}
The lineage of the best program traces 23 ancestry steps from the baseline (program~1) to the best program at roughly program~733 (generation~330), and the single-step jumps along the lineage are large: the first factor-of-three jump occurs between programs~67 and~88 (generations~41 and~50; validation score from \$116K to \$244K), a second between programs~220 and~252 (generations~110 and~121; \$577K to \$904K), and a third between programs~653 and~664 (generations~296 and~302) where the score temporarily regresses before the final push to \$1.85M at program~733. (The lineage scores quoted here are the validation PnL of each ancestor; they sit below the cumulative-best curve plotted in Figure~\ref{fig:run5_progress}.) Unlike the Run~3 joint-strategy evolution, whose per-generation best score occasionally collapsed to near zero before recovering, the Run~5 trajectory is broadly monotonic. A plausible explanation is that the extra axis of variation (features) gives evolution more productive neighbourhoods to explore at any given point, so the optimiser rarely has to pass through deep fitness valleys to find improvements.

\begin{figure}[h]
    \centering
    \includegraphics[width=0.7\textwidth]{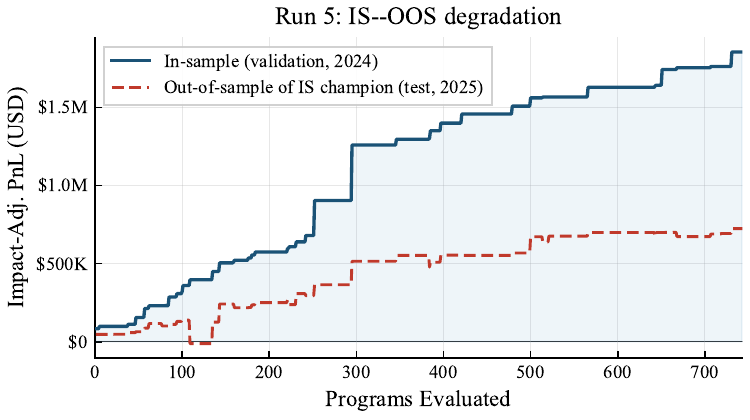}
    \caption{IS--OOS degradation for Run~5. Solid line: best in-sample (validation, 2024) impact-adjusted PnL up to each generation---the quantity evolution optimises. Dashed line: out-of-sample (test, 2025) impact-adjusted PnL of the program currently holding the IS record, i.e., the OOS score one would have obtained by stopping evolution at that generation and deploying the then-current IS champion. The gap between the two curves is the observed IS--OOS haircut; its relation to the multiple-testing baseline of~\citet{bailey2014backtest} is discussed in Section~\ref{sec:discussion}.}
    \label{fig:run5_is_oos_degradation}
\end{figure}

\paragraph{IS--OOS degradation.}
Figure~\ref{fig:run5_is_oos_degradation} resolves the in-sample/out-of-sample gap along the evolutionary trajectory, addressing the iteration-depth question raised in Section~\ref{sec:intro}. The solid line is the best validation (IS) score up to each program---this is the quantity evolution explicitly optimises---while the dashed line reports the \emph{test} (OOS) impact-adjusted PnL of the program that currently holds the IS record. The dashed line therefore answers the question \enquote{if we had stopped evolution at generation $g$ and deployed the then-current IS champion, what would the OOS PnL have been?}; the vertical gap between the two curves is the observed IS--OOS degradation. Two features of the trajectory are worth highlighting. First, the OOS curve is broadly monotone non-decreasing over the run and rises by more than an order of magnitude from roughly \$47K at the baseline to \$724K at the final IS champion, which is qualitatively inconsistent with a pure p-hacking regime in which continued IS selection would leave the OOS curve flat (or send it downward). Second, the short-lived excursion of the OOS curve below zero around program~109 (generation~61) corresponds to an IS champion that overfitted enough to be unprofitable out of sample, but it was displaced at program~135 (generation~71) by a successor whose gains carry across the split. A quantitative comparison against the multiple-testing haircut of~\citet{bailey2014backtest} is deferred to Section~\ref{sec:discussion}.

\begin{figure}[h]
    \centering
    \includegraphics[width=\textwidth]{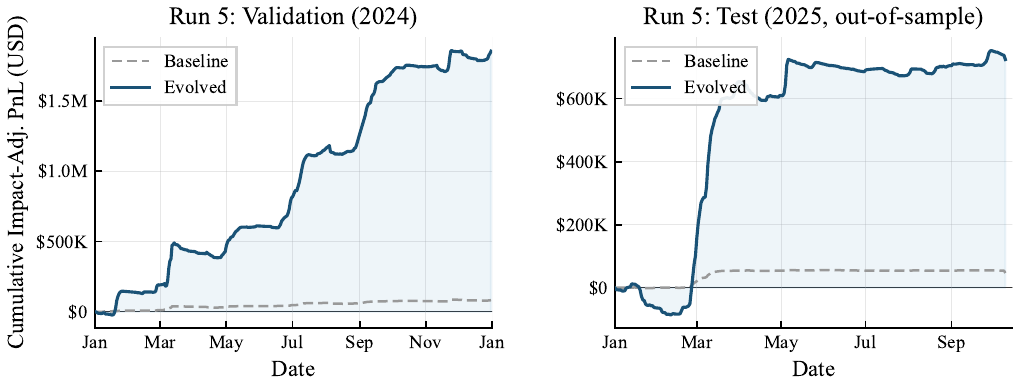}
    \caption{Cumulative impact-adjusted PnL for the baseline and best evolved pipeline in Run~5 (joint evolution of \texttt{default\_calcset} and \texttt{set\_passive\_order\_data}). Left: validation set (2024). Right: test set (2025).}
    \label{fig:run5_pnl}
\end{figure}

\subsection{Hyperparameter Calibration of Algorithms}
\label{sec:hyperparam_calibration}

In this section, we isolate a simpler optimization problem: tuning the parameters of the execution strategy while leaving the pricing and execution logic unchanged. Every trading strategy contains free parameters---for example volatility thresholds, risk-reduction factors, and order-placement depths---and these parameters can materially affect realized PnL even when the underlying alpha model is fixed. Bayesian optimization is a natural tool for this setting because the search space is low-dimensional, the objective is expensive to evaluate, and gradient information is unavailable.

We perform two experiments: (i) calibrating the baseline passive execution strategy when driven by the baseline forecaster, and (ii) recalibrating the same strategy after swapping in the evolved forecaster from Run~4 (Section~\ref{sec:run4}). The configuration parameters of the experiments are given in the Table \ref{tab:hyperparam_setup}. The central question is whether better forecasting metrics automatically translate into better trading performance. As already suggested by the forecasting results, the answer is no: a higher $R^2$ does not necessarily produce higher PnL out of the box. In our experiments, the evolved forecaster does \emph{not} outperform in realized PnL when paired with the default strategy parameters, despite its substantially better predictive metrics. However, after hyperparameter recalibration, it delivers meaningful out-of-sample gains.

\subsubsection{Optimization Method}

We use Optuna's\footnote{https://optuna.org/} Tree-structured Parzen Estimator (TPE), a sequential model-based optimization method well suited to moderate-dimensional hyperparameter search. Unlike Gaussian-process-based approaches, which model the objective function directly, TPE models the distribution of hyperparameters conditioned on observed performance. Specifically, it fits two density estimates over parameter space: $\ell(x)$ for configurations associated with the best trials, and $g(x)$ for the remaining trials. New candidates are then sampled to favor large values of $\ell(x)/g(x)$, which in the TPE framework corresponds to an expected-improvement criterion.

Each sweep begins with 30 uniformly random trials to initialize the density estimators, followed by 90 TPE-guided trials, for a total of 120 trials. We use a fixed random seed for reproducibility. The optimization target is impact-adjusted PnL on the 2024 validation set, defined as total PnL after trading fees and estimated market-impact costs. The 2025 test period is held out entirely and evaluated only once, after the hyperparameter search is complete.

\subsubsection{Experimental Setup}

The strategy under calibration is the base strategy (see Sec. \ref{sec:basestratexplain} and App. \ref{app:basestrat}), which converts an alpha forecast from a Ridge regression model into passive limit orders. 
The trading logic is identical across both experiments; only the forecasting model that feeds the executor changes. 

\begin{table}[t]
\centering
\small
\begin{tabular}{lll}
\toprule
 & \textbf{Experiment (i)} & \textbf{Experiment (ii)} \\
\midrule
Forecasting model & Baseline Ridge (3 features) & Evolved Ridge (77 features) \\
Features & \texttt{ret\_1}, \texttt{ret\_10}, \texttt{ret\_100} & 77 evolved features from \texttt{default\_calcset} \\
$\alpha_{\mathrm{sd}}$ & $3.22 \times 10^{-5}$ & $1.56 \times 10^{-4}$ \\
Data & Polygon BTCUSD, 1-minute bars & Same \\
Training period & 2022--2023 & Same \\
Validation & 2024 (366 days) & Same \\
Test (held-out) & Jan--Oct 2025 (283 days) & Same \\
Trials & 120 (30 random + 90 TPE) & Same \\
\bottomrule
\end{tabular}
\caption{Hyperparameter-calibration setup for the two forecaster variants.}
\label{tab:hyperparam_setup}
\end{table}

In both experiments, we tune the same eight strategy parameters using identical bounds and scaling (see Appendix~\ref{app:hyperparam_details}). The capital constraint $q_{\max}=\$200{,}000$ (see App. \ref{app:basestrat}) is fixed and is not optimized.

\subsubsection{Results}

\begin{table}[t]
\centering
\small
\setlength{\tabcolsep}{5pt}
\renewcommand{\arraystretch}{1.05}
\begin{tabular}{llcc}
\toprule
\textbf{Configuration} & \textbf{Metric} & \textbf{Baseline forecaster} & \textbf{Evolved forecaster} \\
\midrule
\multicolumn{4}{l}{\textit{Validation --- default parameters}} \\
 & Impact-adjusted PnL & \$82{,}615 & \$206{,}898 (2.50x) \\
 & Sharpe & 4.81 & 7.65 (1.59x) \\
 & Max drawdown & -\$8{,}739 & -\$7{,}391 (0.85x) \\
\midrule
\multicolumn{4}{l}{\textit{Test --- default parameters}} \\
 & Impact-adjusted PnL & \$46{,}791 & \$27{,}842 (0.59x) \\
 & Sharpe & 3.82 & 2.47 (0.65x) \\
 & Max drawdown & -\$8{,}200 & -\$49{,}053 (5.98x) \\
\midrule
\multicolumn{4}{l}{\textit{Validation --- calibrated parameters}} \\
 & Impact-adjusted PnL & \$225{,}180 & \$730{,}783 (3.25x) \\
 & Sharpe & 8.56 & 9.35 (1.09x) \\
 & Max drawdown & -\$6{,}774 & -\$13{,}943 (2.06x) \\
 & Win rate & 77.9\% & 65.6\% (0.84x) \\
\midrule
\multicolumn{4}{l}{\textit{Test --- calibrated parameters}} \\
 & Impact-adjusted PnL & \$103{,}089 & \$159{,}967 (1.55x) \\
 & Sharpe & 4.15 & 4.15 (1.00x) \\
 & Max drawdown & -\$16{,}782 & -\$87{,}866 (5.24x) \\
 & Win rate & 65.4\% & 36.0\% (0.55x) \\
\bottomrule
\end{tabular}
\caption{Hyperparameter-calibration results for the baseline and evolved forecasters before and after Optuna tuning. Values in parentheses in the evolved-forecaster column report the ratio relative to the baseline forecaster within the same block.}
\label{tab:hyperparam_results}
\end{table}

\paragraph{PnL before and after calibration.} Table~\ref{tab:hyperparam_results} combines the default and calibrated results for both forecasters on the validation and test splits. With default parameters, the evolved forecaster appears much stronger in-sample: it generates roughly $2.5\times$ the validation PnL of the baseline forecaster (Table~\ref{tab:hyperparam_results}). But this gain does not survive out of sample. On the held-out 2025 test period, the evolved forecaster underperforms the baseline (\$27{,}842 versus \$46{,}791) and experiences a much larger drawdown. 

Calibration produces large gains in both experiments. For the baseline forecaster, test impact-adjusted PnL increases from \$46{,}791 to \$103{,}089, a gain of about 120\%, while the Sharpe ratio improves from 3.82 to 4.15. For the evolved forecaster, the effect is even larger: test impact-adjusted PnL rises from \$27{,}842 to \$159{,}967, a roughly 475\% increase. In other words, the evolved forecaster goes from the worst-performing configuration out of the box to the best-performing one after recalibration.

These results show that hyperparameter recalibration is not optional when changing forecasting models. The default strategy parameters are implicitly tuned to the scale and signal-to-noise characteristics of the baseline alpha. Once the alpha model changes, position sizing, trade filtering, and risk management must be re-optimized as well; otherwise, the stronger forecaster can be made to look worse simply because the executor is miscalibrated. In absolute terms, the calibrated evolved model exceeds the calibrated baseline by roughly \$57{,}000 on the test set, which is the economically relevant comparison.

\begin{figure}[t]
    \centering
    \begin{minipage}[t]{0.49\textwidth}
        \centering
        \includegraphics[width=\linewidth]{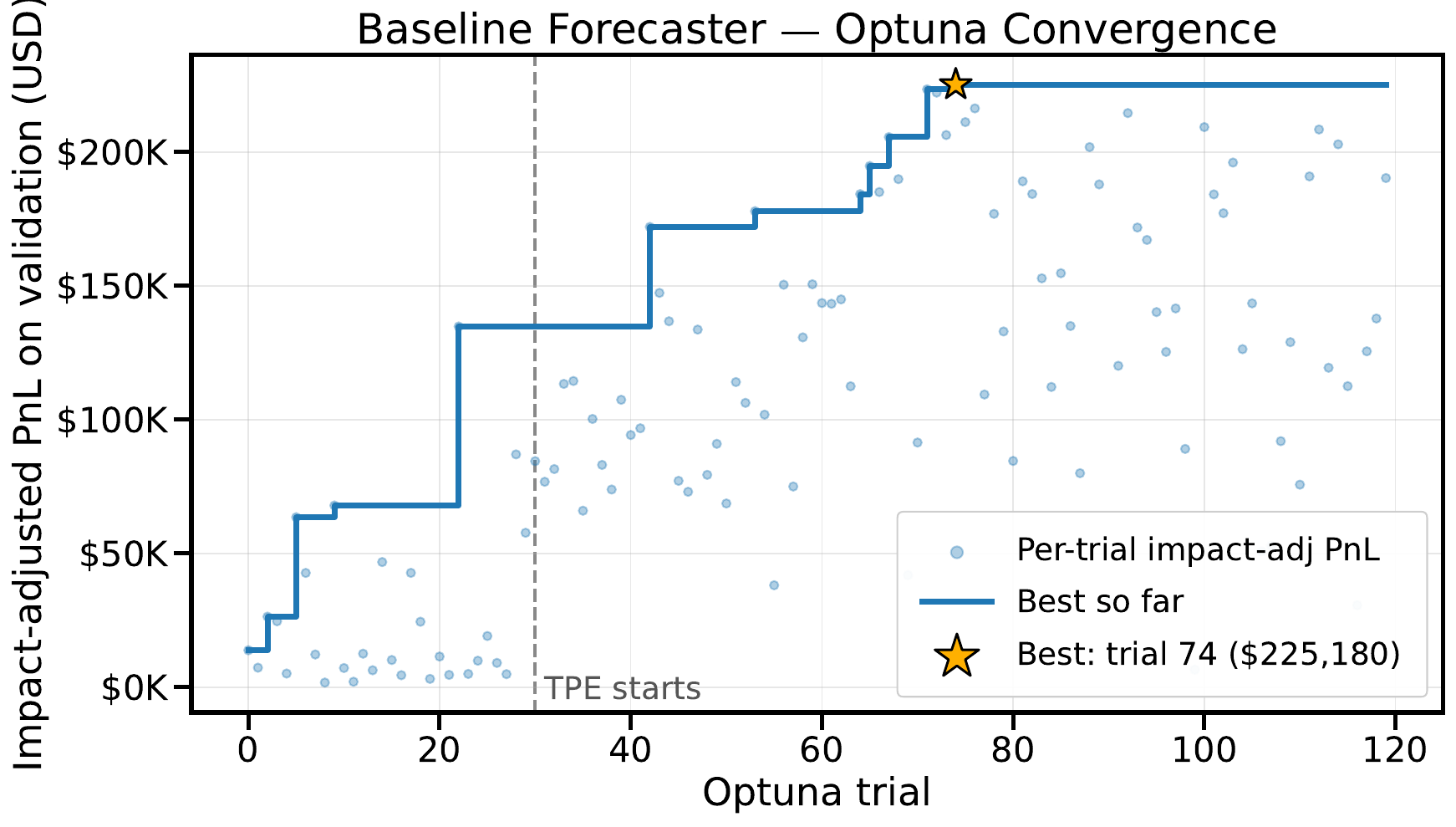}
    \end{minipage}\hfill
    \begin{minipage}[t]{0.49\textwidth}
        \centering
        \includegraphics[width=\linewidth]{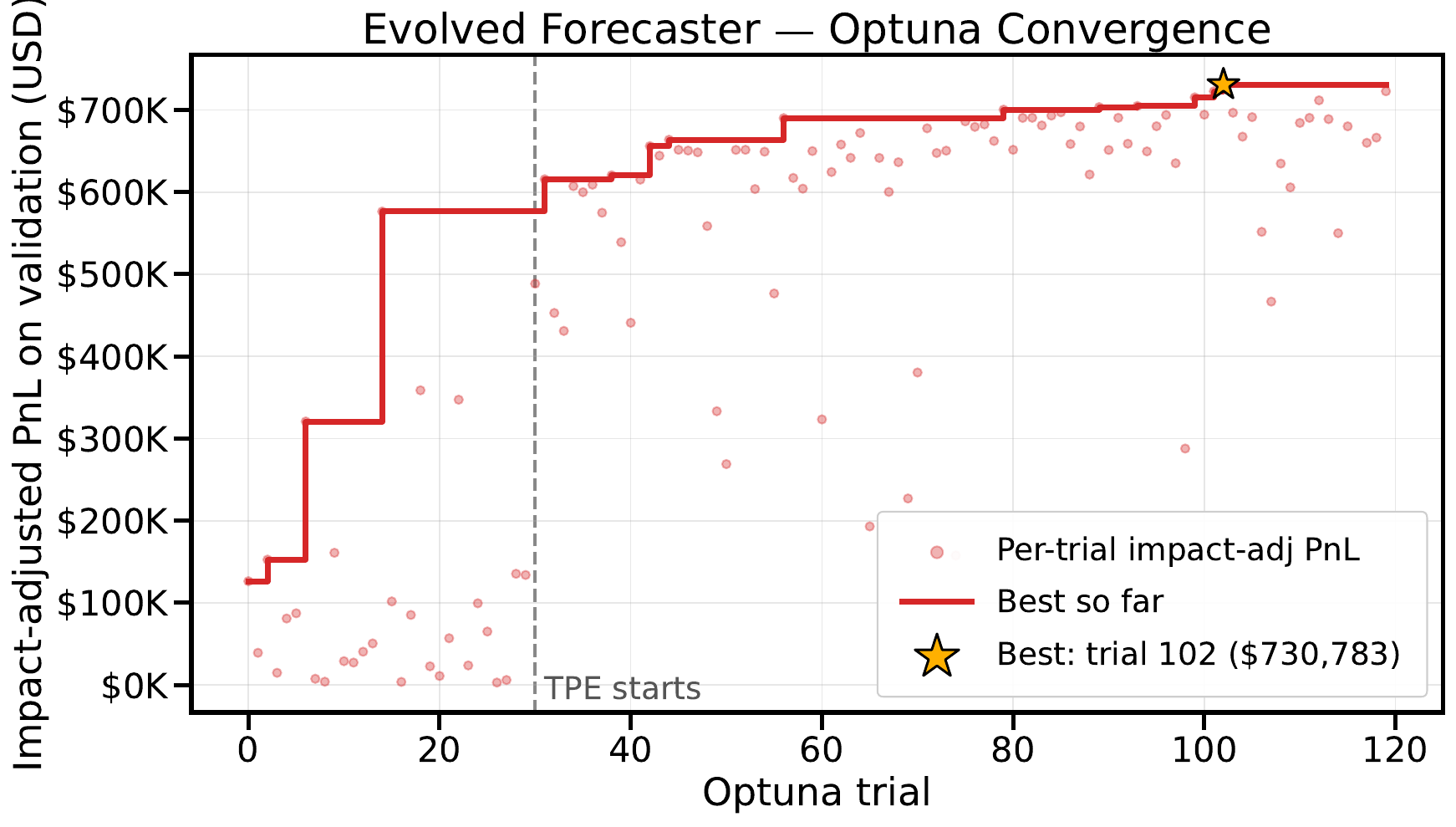}
    \end{minipage}
    \caption{Optuna convergence diagnostics for the two hyperparameter sweeps. Left: baseline forecaster. Right: evolved forecaster. Each plot shows per-trial validation impact-adjusted PnL together with the running best across 120 trials.}
    \label{fig:hyperparam_convergence}
\end{figure}

\paragraph{Optimization Convergence.} The per-trial validation scores suggest a similar qualitative picture in both sweeps: roughly half of the total improvement is found during the 30 random warm-up trials, and TPE then refines the search from there. The evolved-forecaster sweep reaches a much higher validation plateau (\$731{,}000 versus \$225{,}000) but appears to converge more slowly, consistent with the richer alpha giving the optimizer a larger productive region to explore. We visualize the Optuna optimization in Fig. \ref{fig:hyperparam_convergence}.

\paragraph{Tuned Parameters.}

Table~\ref{tab:hyperparam_params} in Appendix~\ref{app:hyperparam_details} lists the eight effective strategy parameters, their defaults, search bounds, and the best values found in the two sweeps. Several differences between the two optima are informative. First, the baseline sweep pushes \texttt{sizing\_factor} close to the upper bound, compensating for the weaker alpha by taking larger positions, whereas the evolved sweep chooses a more moderate value, likely because the signal is stronger. Second, the two sweeps choose very different entry depths \texttt{zp}: the baseline forecaster prefers wider passive entries, while the evolved forecaster prefers nearly at-the-mid placements, consistent with a stronger signal that decays quickly. Third, the evolved optimum almost disables the \texttt{alpha\_adjustment\_knob}, suggesting that once the forecaster is strong enough, the strategy benefits from carrying more conviction rather than aggressively penalizing inventory.

\subsection{Model Ensemble Analysis}
\label{sec:model_analysis}

\begin{figure}[h]
    \centering
    \includegraphics[width=\textwidth]{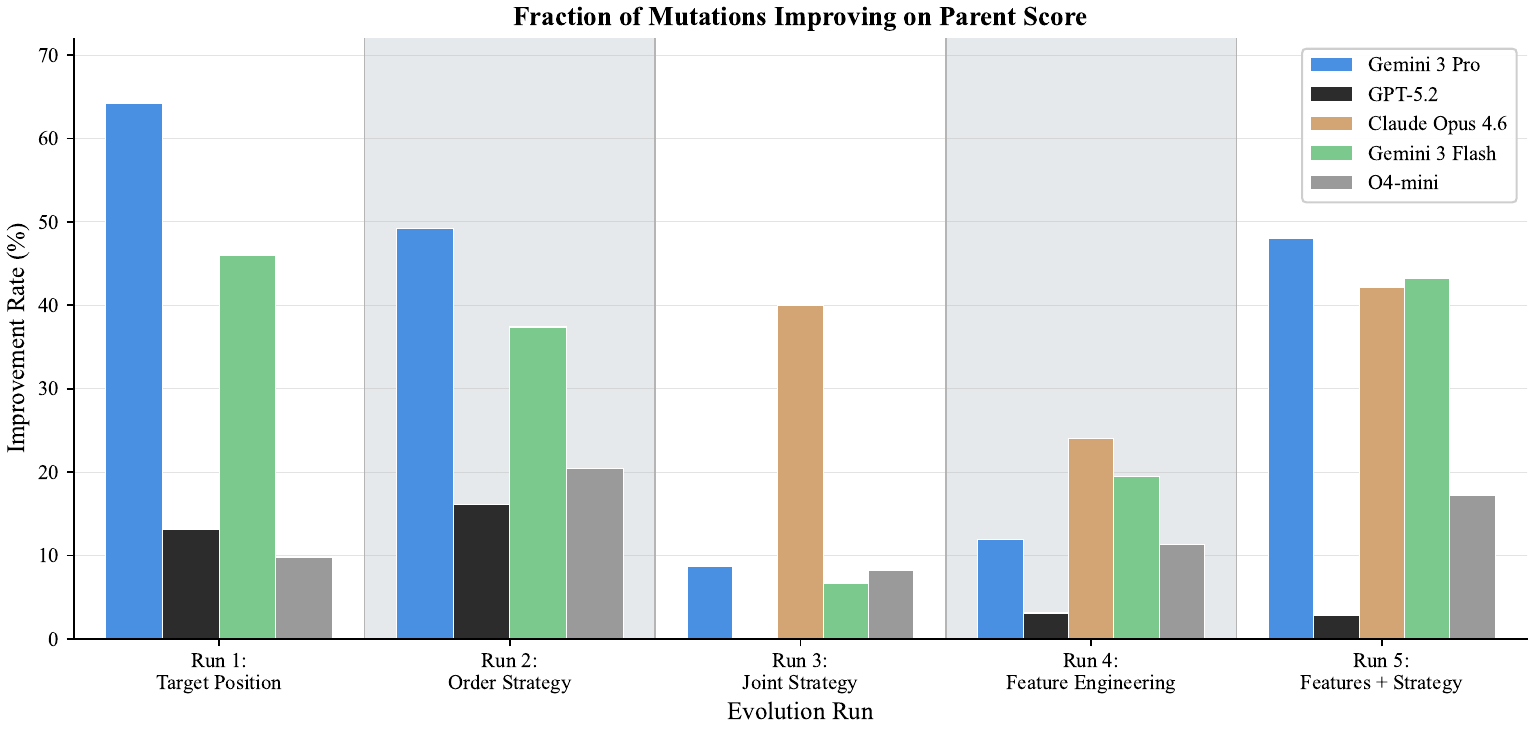}
    \caption{Per-model improvement rates (fraction of mutations that exceed the parent's fitness) across the five evolution runs. Rates are highest on the component-wise strategy runs (Runs~1--2), drop sharply on the harder Runs~3 and~4, and partially rebound on Run~5. Claude Opus~4.6 was only included in the ensemble for Runs~4 and~5, so its bars are absent for Runs~1--3.}
    \label{fig:model_improvement}
\end{figure}

As described in Section~\ref{sec:framework_llm}, \textsc{MadEvolve} routes mutation queries across an ensemble of frontier LLMs. All five evolution runs reported above used a pool of four to five models: Gemini~3 Pro, Gemini~3 Flash, GPT-5.2, o4-mini, and Claude Opus~4.6. Each model received a roughly equal share of mutation requests, allowing a controlled comparison of how different models contribute to the evolutionary process. Table~\ref{tab:model_stats} summarizes aggregate performance across all 4{,}559 evolved programs, and Figure~\ref{fig:model_improvement} shows per-run improvement rates.

\begin{table}[h]
\centering
\small
\begin{tabular}{llrrrr}
\toprule
\textbf{Model} & \textbf{Tier} & \textbf{Programs} & \textbf{Impr.\ rate} & \textbf{Top-20} & \textbf{Lineage} \\
\midrule
Gemini 3 Pro   & frontier  & 1{,}061 & 40.2\% & 29 & 37 \\
Claude Opus 4.6 & frontier &   297 & 33.0\% & 11 & 13 \\
Gemini 3 Flash & efficient & 1{,}084 & 30.9\% & 40 & 30 \\
O4-mini        & efficient & 1{,}055 & 12.6\% &  9 & 13 \\
GPT-5.2        & frontier  & 1{,}062 &  8.2\% & 11 & 12 \\
\bottomrule
\end{tabular}
\caption{Aggregate model performance across all five evolution runs. \emph{Impr.\ rate}: fraction of mutations that improved on the parent's fitness score. \emph{Top-20}: number of appearances among the 20 highest-scoring programs per run (100 slots total). \emph{Lineage}: number of steps in the ancestry chains of the five best-in-run solutions.}
\label{tab:model_stats}
\end{table}

\paragraph{Improvement rates.} The fraction of mutations that improve on the parent score varies sharply across models and tasks (Fig.~\ref{fig:model_improvement}). Claude Opus~4.6 was only included in the ensemble for Runs~4 and~5, so its statistics for the strategy runs are absent rather than zero; the aggregate counts in Table~\ref{tab:model_stats} reflect this smaller sample. On the component-wise strategy runs (Runs~1--2), the two Gemini variants are clearly the most productive, while GPT-5.2 and o4-mini improve less often but still produce some high-scoring outliers; the best program in Run~2 came from GPT-5.2. Run~3 (joint strategy) sees improvement rates drop sharply, with Gemini~3 Pro the only model holding above ten percent (22\%) and the other three falling to single digits, consistent with the rugged fitness landscape discussed in Section~\ref{sec:exp_joint}. Run~4 (feature engineering) is the only run in which Claude Opus~4.6 has the highest improvement rate. Run~5 (joint feature+strategy) shows a partial rebound for the top three models, while o4-mini and GPT-5.2 trail well behind. The rebound is notable because Run~5 has a strictly larger search space than Run~3; this suggests that the extra variation axis supplied by the feature pipeline opens up more productive neighbourhoods to explore, even as the dimensionality grows.

\paragraph{Top solutions and lineage.} While improvement rate measures average mutation quality, the top-scoring programs and the lineages of the best solutions reveal a complementary picture. Here, a lineage is the chain of parent programs that links a given solution back to the baseline through successive mutations. Gemini~3 Flash contributes the most top-20 programs (40 of 100 slots) and 30 of 105 lineage steps. At first sight this is counterintuitive, since Flash has the lower average improvement rate of the two Gemini variants. A natural explanation is that Flash's mutations have higher variance: more outright failures, but also a longer tail of large successes, which is what the top-20 metric picks up. Gemini~3 Pro contributes 37 lineage steps, the single largest share, confirming that its high improvement rate translates into sustained evolutionary progress. Claude Opus~4.6, although only available in Runs~4--5, still contributes 13 lineage steps. The best program in Run~5 was itself produced by Gemini~3 Flash, and its 23-step lineage draws on Gemini~3 Pro (8 steps), Claude Opus~4.6 (7 steps), Gemini~3 Flash (6 steps), and o4-mini (2 steps), with no contribution from GPT-5.2. More broadly, no single model dominates the lineage of any best solution: the best programs in all five runs trace their lineage through three to five distinct models, with alternating contributions from frontier and efficient models. This confirms that the ensemble's value lies not in any individual model's superiority but in the diversity of mutation strategies they collectively provide.

\paragraph{Takeaway.} Two caveats apply to the rankings reported here. First, the conventional frontier-versus-efficient labeling does not cleanly predict contribution: GPT-5.2 is the weakest contributor on every metric despite being a frontier model, while Gemini~3 Flash, an efficient-tier model, produces the most top-20 programs. Second, the specific rankings will date quickly. Frontier model releases happen on a timescale of weeks, and re-running these experiments a few months from now would almost certainly show a different distribution of contributors. The qualitative finding is what we expect to persist: a diverse ensemble outperforms any single model in our setup.

\section{Comparison with Claude Code}
\label{sec:claudecode}

In principle, a wide variety of agentic approaches can be used to optimize the target metrics in this paper, once a reliable simulation setup is developed. We thus carried out exploratory research into an alternative algorithmic search approach, for comparison with the evolutionary scaffolding in our main analysis. We instructed Claude Code (Sonnet 4.6 is being used as the main model) to start from the same initial strategy and iteratively suggest improvements, maintaining the list of ideas and evaluation feedback.

\subsection{Strategy Search}
\label{sec:claude_strategy_search}
We describe the exact workflow in Algorithm~\ref{alg:workflow}, which we implemented as a python script that builds and executes an LLM prompt to Claude Sonnet 4.6 at each iteration. 

\begin{algorithm}[H]
\label{alg:workflow}
\caption{LLM-Guided Trading Strategy Evolution with Claude}
\begin{algorithmic}[1]

\Require Template strategy with two mutable code components, backtester, $N$ iterations
\Ensure Best strategy assembled from best version of each component

\State Evaluate baseline strategy $\to$ initial score and feedback
\State Initialize ideas tree with baseline result

\For{$i = 1, \dots, N$}

    \State Build prompt containing:
    \Statex \hspace{\algorithmicindent} task description
    \Statex \hspace{\algorithmicindent} ideas tree summary (all past mutations, scores, lineage)
    \Statex \hspace{\algorithmicindent} current best code for both components
    \Statex \hspace{\algorithmicindent} metrics from previous iteration

    \State Query LLM $\to$ returns which component to modify, new code, reasoning

    \State Assemble strategy: new code for modified component + best code for the other
    \State Run backtest on validation data
    \State $\text{score} \gets \text{PnL} - \text{MarketImpactCost}$

    \State Record result in ideas tree
    \If{score improved}
        \State Update best code for that component
    \EndIf
    \State Store evaluation metrics as feedback for next iteration

\EndFor

\end{algorithmic}
\end{algorithm}

We made the following design choices to help Claude Code explore the space of possibilities:
\begin{itemize}
\item \textbf{Dual-component evolution.}
The strategy has two independently evolvable parts: \texttt{set\_target()} decides \emph{what} to trade
(position sizing, risk limits, trade filters), and \texttt{set\_limit\_order()} decides \emph{how} to
price orders. The LLM chooses which part to modify each iteration, allowing focused optimization
of one concern at a time.
\item \textbf{Ideas tree as memory.}
Every mutation is recorded in a tree with its score, parent idea, and evaluation metrics.
This tree is included in every prompt, so the LLM can see the full history of what worked
and what failed---enabling it to refine winners, avoid repeating failures, and backtrack
to promising earlier branches. 
\item \textbf{Feedback loop.}
After each evaluation, the LLM receives quantitative feedback: Sharpe ratio, PnL,
maximum drawdown, win rate, and market impact cost, however, we instruct the model to optimize for impact-adjusted PnL.
\item \textbf{Backtester.}
Each candidate is evaluated via minute-bar simulation, using the same backtesting engine and time splits as in the rest of this work.
\end{itemize}
We ran this procedure for 200 iterations using Claude Sonnet 4.6 as the underlying model and a 10-minute timeout per backtest. Out of 200 proposed mutations, 170 were evaluated successfully, while 30 failed because of syntax errors, malformed outputs, or timeouts. Throughout this experiment, the alpha model was kept fixed to the baseline Ridge forecaster with three input features, as defined in App. \ref{ssec:baseline_alpha}, so the search operated entirely over execution and sizing logic rather than forecasting.

The baseline strategy achieves \$82{,}615 of impact-adjusted PnL on the 2024 validation set. After 200 iterations, the best Claude-discovered strategy reaches \$583{,}783, corresponding to a $7.1\times$ improvement. On the held-out January--October 2025 test set, the same strategy achieves \$340{,}326 of impact-adjusted PnL, retaining about 58\% of its validation performance. Impact costs are stable across the two splits at roughly 0.9 basis points of traded volume

\begin{table}[t]
\centering
\small
\begin{tabular}{lcc}
\toprule
 & \textbf{Validation (2024)} & \textbf{Test (Jan--Oct 2025)} \\
\midrule
PnL after fees & \$1{,}013{,}302 & \$545{,}076 \\
Impact cost & \$429{,}519 & \$204{,}750 \\
Impact-adjusted PnL & \$583{,}783 & \$340{,}326 \\
Sharpe & 4.91 & 3.54 \\
Sortino & 14.99 & 5.38 \\
Max drawdown & -\$79{,}512 & -\$79{,}164 \\
Win rate & 61.5\% & 55.5\% \\
Days & 366 & 283 \\
Volume traded & \$4.66B & \$2.23B \\
Number of trades & 306{,}304 & 164{,}313 \\
\bottomrule
\end{tabular}
\caption{Best strategy found by Claude Code in the strategy-search experiment.}
\label{tab:claude_strategy_search}
\end{table}

The search exhibits three broad phases. In iterations 1--28, the score improves rapidly from roughly \$82k to \$194k, driven mainly by changes to \texttt{set\_target()} such as trade-size filtering and position-scaling adjustments. Between iterations 28 and 110, the search makes a second large jump to about \$460k after discovering a convex power-law sizing rule that concentrates capital on high-conviction signals and reduces turnover on weaker ones. From iteration 110 onward, progress becomes incremental, eventually reaching \$584k by refining the sizing logic and the limit-order placement policy.

We also find a pronounced asymmetry in where the gains come from. Across the 170 successful evaluations, Claude modified the target-position component in 104 iterations (61\%), the limit-order component in 54 iterations (32\%), and both simultaneously in 12 iterations (7\%). This suggests that, in this setup, most of the available edge comes from improving \emph{what} to trade rather than \emph{how} to price orders. The final strategy reflects this: compared with the baseline, it introduces convex power-law position sizing, momentum-adaptive order depth, and a more asymmetric risk-off schedule for unwinding losing positions. 

\subsection{Feature Evolution}

We also applied the same LLM-guided search framework to the forecasting pipeline rather than the trading policy itself. In this experiment, Claude Code evolves the body of \texttt{default\_calcset()}, the function that maps raw OHLCV minute-bar data to the feature matrix used by the Ridge forecaster. The procedure mirrors the strategy-search workflow, but differs in both the mutable object and the evaluation pipeline.

The feature-search loop is:
\begin{enumerate}
    \item Claude receives the current best \texttt{default\_calcset()}, the ideas tree summarizing previous proposals and scores, and the feedback from the last successful evaluation.
    \item Claude proposes a new calcset function that may add, remove, or restructure candidate features.
    \item The evaluator computes \emph{all} candidate features generated by that calcset on the 2022--2023 training split.
    \item The evaluator ranks all candidate features by their univariate $|\mathrm{corr}|$ with the 10-minute forward return.
    \item Features are then selected greedily in that rank order, with a pairwise-correlation filter: a candidate is kept only if its absolute correlation with \emph{every} already-selected feature is less than 0.85. This enforces approximate orthogonality and prevents the model from filling the design matrix with near-duplicates.
    \item The greedy selection stops once no admissible features remain or once 10 features have been selected.
    \item A Ridge model with $\alpha=0.5$ is fit on the selected features, and the resulting alpha signal is evaluated on the 2024 validation set.
    \item The candidate receives the score ($R^2$ in our case), and the ideas tree is updated. If the score improves on the incumbent, the proposed calcset becomes the new best program for the next iteration.
\end{enumerate}

We ran 200 Claude iterations with a 5-minute timeout per call, with the target metric being the forecasting $R^2$ of a log return at a 10-minute time horizon. 
The baseline forecasting model uses three hand-crafted momentum features, while the best evolved one proposes 23 candidate features, of which 10 survive forward selection. On the validation split, the evolved forecaster improves $R^2$ from 0.0021 to 0.0105.

On the held-out 2025 test set, the gains persist: $R^2$ rises from 0.0017 to 0.0091. We note that here we were able to achieve even higher values of $R^2$ than in Sec. \ref{sec:run4}, which might be peculiar to the aggregated candlestick data used.

\begin{table}[t]
\centering
\small
\setlength{\tabcolsep}{5pt}
\renewcommand{\arraystretch}{1.05}
\begin{tabular}{lcc}
\toprule
\textbf{Metric} & \textbf{Baseline} & \textbf{Evolved} \\
\midrule
\multicolumn{3}{l}{\textit{Validation (2024)}} \\
$R^2$ & 0.0021 & 0.0105 (5.00x) \\
Correlation & 0.077 & 0.105 (1.36x) \\
\midrule
\multicolumn{3}{l}{\textit{Test (2025)}} \\
$R^2$ & 0.0017 & 0.0091 (5.35x) \\
Correlation & 0.064 & 0.097 (1.52x) \\
\bottomrule
\end{tabular}
\caption{Forecast-quality metrics for the baseline and Claude-evolved calcsets. Values in parentheses in the evolved column report the ratio relative to baseline.}
\label{tab:claude_feature_evolution}
\end{table}

The selected features indicate that Claude does not simply rediscover the original momentum basis. The evolved calcset keeps the core short-horizon returns (such as \texttt{ret\_1} and \texttt{ret\_5}) but adds three new signal families: lagged-return terms at multiple offsets, VWAP-derived 
features, and volatility-return interaction terms. Qualitatively, this broadens the forecaster from a pure momentum model into a richer microstructure-aware representation. 

Improved alpha quality, however, does not automatically translate into improved trading PnL. When plugged into the default execution strategy, the evolved forecaster actually performs worse out of sample: test impact-adjusted PnL falls from \$46{,}791 for the baseline alpha to -\$27{,}191 for the evolved alpha, with max drawdown widening from -\$8{,}200 to -\$54{,}420. The reason is that the evolved alpha has a much larger scale ($\alpha_{\mathrm{sd}}$ increases by about $5.4\times$), so the default position-sizing and order-placement constants are no longer calibrated correctly.

After re-tuning the same eight execution hyperparameters with Optuna (as described in Sec. \ref{sec:hyperparam_calibration}), the evolved forecaster's advantage becomes economically visible. The recalibrated baseline reaches \$103{,}089 of test impact-adjusted PnL, whereas the recalibrated evolved alpha reaches \$235{,}419 with a Sharpe ratio of 5.27 and max drawdown of -\$20{,}146. In other words, the Claude-evolved features do appear to add genuine predictive signal, but extracting that signal requires a second calibration stage in the execution layer.

\subsection{Observations}

We report the following observations based on the results of our experiments. The presented approach has proven to be promising - unlike random mutations in the evolutionary search, the changes suggested by maintaining the idea tree make exploration more systematic and complete. However, we noticed that the results are noisy and sensitive to the exact prompt and the workflow instructions. For example, in the run reported here, we were able to produce a strategy that is comparable in performance to the one obtained with MadEvolve. In another run, maintaining the same number of iterations and a slightly modified prompt, we obtained only minimal improvements (we achieved only 48.8\% and 44.2\% improvement in IS and OOS PnL correspondingly, as opposed to 607\% and 627\% for the run described in Sec. \ref{sec:claude_strategy_search}).

The evolutionary scaffolding in the main analysis may keep algorithms leaner and more efficient, as only the best-performing mutations are kept. However, Claude Code can be prompted to perform similarly (e.g. to not exceed a certain number of code lines or free parameters). Given the prompt sensitivity and stochasticity of the optimization, we cannot unambiguously conclude which particular framework performs better, and for practical use one might try a number of approaches in parallel on the same target metrics, if sufficient compute is available. 

Claude feature search for the forecasting model has proven to be more successful than strategy search. After parameter recalibration, the baseline strategy with claude-evolved forecaster achieves \$235{,}419 impact-adjusted PnL with a Sharpe of 5.27 on the test set (as compared to \$159{,}967 and Sharpe ratio of 4.15 with MadEvolve). Additionally, claude-evolved features maintain better consistency in IS vs OOS performance. However, the gains might be partially attributed to regularization techniques that we used during the search with Claude such as a constraint on a number of features and their pairwise correlation. We note that there might be a better approach for maintaining an uncorrelated feature set. We defer this exploration to future work


\section{Research or P-Hacking? The Central Question.}
\label{sec:discussion}

In this section, we investigate in more detail to what extent the gains obtained with our evolutionary framework can be explained by chance due to the number of trials, otherwise known as look-elsewhere effect or p-hacking. First, let us illustrate why the backtest hacking occurs. We will demonstrate this on the example of Sharpe ratio for simplicity, although a similar argument holds for any metrics used for the strategy selection. As was explicitly shown in \cite{bailey2014backtest}, maximum Sharpe ratio grows as $\mathcal{O}(\sqrt{\ln N})$ with $N$ independent trials, and therefore it's possible to obtain an unprofitable strategy with a high Sharpe ratio in the backtest due to a large trial factor. We will argue here that this is not the case for the strategies obtained with algorithmic evolution, as described in this work. 

\begin{figure}[h]
    \centering
    \includegraphics[width=0.9\textwidth]{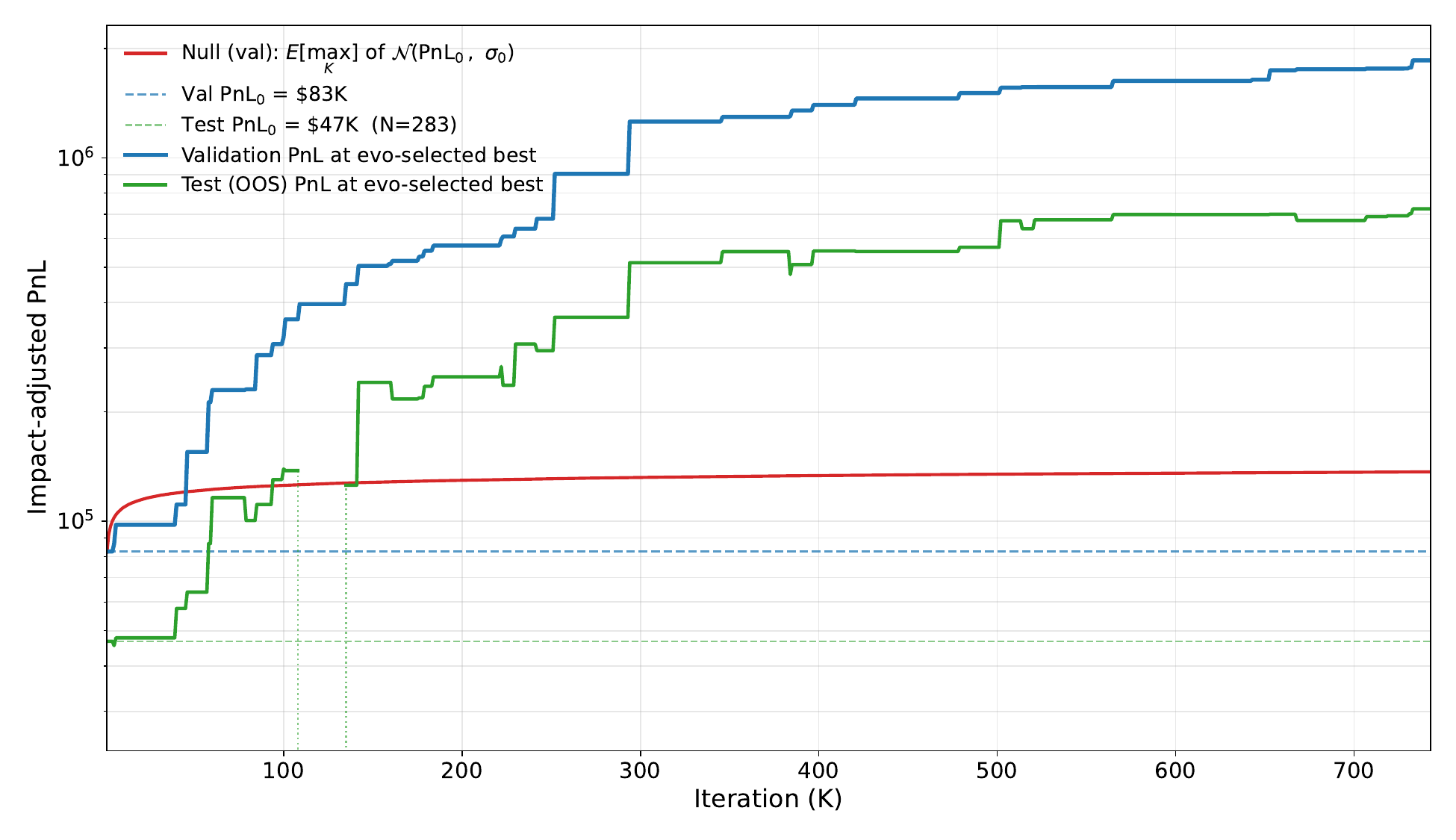}
    \caption{Null test against a baseline-shifted $p$-hacking procedure.
    Blue: validation impact-adjusted PnL at the evolution-selected best.
    Green: held-out test PnL of the same evo-selected strategy.
    Red: expected best-of-$K$ if each iteration's PnL were drawn from
    $\mathcal{N}(\mathrm{PnL}_0,\,\sigma_0)$ on validation --- the
    hypothetical $p$-hacking ceiling. Dashed lines mark the baseline
    $\mathrm{PnL}_0$ for each split; the test $\sigma_0$ is the validation
    $\sigma_0$ rescaled by $\sqrt{283/366}$ to account for the shorter
    test window. Vertical dotted green segments mark iterations where the
    test PnL briefly dipped below zero and is hidden by the logarithmic
    axis.
    }
    \label{fig:null_comparison}
\end{figure}

Consider $\{X_n\} = \{X_1,...,X_n\}, X\sim N(0,1)$ - a sequence of i.i.d. random normal variables of length N, and $Y_N = \max \{X_n\}$. Then a standard result is (e.g. \cite{bailey2014backtest} and references therein) that $\mathbb{E}[Y_N] \leq \sqrt{2\ln N}$. In other words, it is expected to find strategies with progressively higher Sharpe ratio by simply trying more 
strategies with from a distribution with mean zero excess return. We are able to convince ourselves that our results are not a p-hacking due to the following two reasons:
\begin{itemize}
    \item The main argument is that in the case of p-hacking, one should not expect the performance gains to translate to OOS data. We used a hold-out set (test-set) of data to evaluate the OOS performance and confirmed that the gains do generalize. 
    \item Secondly, even for the in-sample results alone, for the typical amount of trials in our experiment, the obtained Sharpe ratios are unlikely to be random flukes as we quantify below. Moreover, the assumption of independence in the null hypothesis is usually conservative because in the case of perfect correlation $X_1=...=X_n$, $\mathbb{E}[Y_N]$ doesn't grow with $N$. In our case, evolutionary strategy search is not independent - each subsequent iteration relies on the previous history. Therefore, we should not expect to obtain as high Sharpe ratios even on the data used for evolution. 
\end{itemize}
Figure~\ref{fig:null_comparison} clearly demonstrates these arguments quantitatively. We
choose the evolutionary run of the trading strategy joint with features, described in
Sec.~\ref{sec:exp_joint_calcset_strategy}, for illustration. The solid blue
line shows the impact-adjusted PnL of the evolution-selected best strategy
on the validation set as a function of iteration; the solid green line
shows the PnL of that same strategy on the held-out test set (which was
never used for selection). The evolution optimised the validation PnL, so
the blue curve is monotone by construction, while the green curve inherits
this selection only indirectly and is free to regress under a truly
overfit procedure --- but here it does not.

Instead of testing against the classical ``the strategy has no edge''
null, we benchmark against a much weaker, more conservative alternative:
a hypothetical $p$-hacking procedure that simply draws each iteration's
PnL from a Gaussian $\mathcal{N}(\mathrm{PnL}_0,\,\sigma_0)$ centred on
the baseline's observed performance, and reports the best-of-$K$. The
solid red line is the expected ceiling of such a procedure,
$\mathrm{PnL}_0 + \sigma_0\sqrt{2\ln K}$. We calibrate $\sigma_0$ directly
from the baseline: $\sigma_0 = \mathrm{PnL}_0/S_0$, where $S_0$ is the
baseline's annualised Sharpe and $\mathrm{PnL}_0$ its total
impact-adjusted PnL. For the validation window
($N_\mathrm{val}\!=\!366$, $S_0\!=\!4.81$,
$\mathrm{PnL}_0\!=\!\$82.6\mathrm{K}$) this yields
$\sigma_0^\mathrm{val}\!\approx\!\$17.2\mathrm{K}$. Because the test
window is shorter ($N_\mathrm{test}\!=\!283$ days), its null standard
deviation is rescaled by $\sqrt{283/366}$, giving
$\sigma_0^\mathrm{test}\!\approx\!\$15.1\mathrm{K}$ about the test
baseline $\mathrm{PnL}_0^\mathrm{test}\!=\!\$46.8\mathrm{K}$.

\textit{The final evolution-selected strategy reaches
$\mathrm{PnL}=\$1.85\mathrm{M}$ on validation and
$\mathrm{PnL}=\$724\mathrm{K}$ on the held-out test set. Expressed in
$z$-units of the baseline null, the validation excess sits at
$z_\mathrm{val}=103.2$ above $\mathrm{PnL}_0^\mathrm{val}$, and the test
excess at $z_\mathrm{test}=44.9$ above $\mathrm{PnL}_0^\mathrm{test}$.
Even accounting for the best-of-$K\!=\!335$ selection on the validation
side, such deviations are so far into the Gaussian tail that a pure
p-hacking procedure around the baseline has essentially no chance of
producing them: both the validation and the test outcomes are
overwhelmingly inconsistent with the baseline-noise null.}

This discussion does not address the question of \textit{backtest overfitting}, in the sense that, while we use an independent OOS test data time range, the backtest logic for both training and test data is the same, and the market regime could change outside of our training and test data periods. In all our experiments, we made certain implicit and explicit assumptions about the \textit{model} of the realized PnL - i.e.
\begin{enumerate}
    \item Forecasting model represents the reality - i.e. the data used for backtest contains a true signal, realizable on the exchange. As discussed above, our backtesting data is exchange aggregated and not directly tradable.
    \item Full fill at limit price.
    \item Certain structure of the fees and market impact model.
\end{enumerate}
We demonstrated that given these assumptions, we are able to improve the strategy performance with the evolutionary search framework. Additionally, as discussed in this section, we find that the improvements are due to genuine financial findings, and not simply due to a large number of experiments. However, due to the assumptions made above, our results do not directly transfer to live trading.

\section{Conclusion}
\label{sec:conclusion}
In this paper, we applied \textsc{MadEvolve}, a general-purpose LLM-driven evolutionary search framework inspired by AlphaEvolve, to a sequence of quantitative-finance problems ranging from target-position design and order placement to feature engineering and full joint feature+strategy optimization. The main result is that agentic evolutionary search can produce large, economically meaningful improvements even in a noisy domain like trading. Across all four PnL-optimizing runs, out-of-sample Sharpe improves by 0.6--1.8 points over the corresponding baselines. The strongest absolute out-of-sample result comes from evolving order placement alone, which raises test impact-adjusted PnL from \$47K to \$1.20M. The best risk-adjusted performance comes from joint feature+strategy evolution, which achieves a test Sharpe of 5.65 and increases test impact-adjusted PnL from \$47K to \$724K. In the forecasting-only setting, the evolved feature set approximately doubles out-of-sample $R^2$, showing that agentic search remains useful even when the objective is substantially noisier than execution optimization. The headline results of the strategy evolution are summarized in Table~\ref{tab:summary}. We note, however, that the demonstrated gains should be interpreted conditionally on the market model assumed by the backtest engine. While we included realistic aspects of it such as the fee structure and market impact model calibrated to the specific exchange, assessment of the live-trading performance is left to future exploration.

Just as importantly, our results are not only about performance but also about methodology. Finance is precisely the setting where a powerful search loop can collapse into p-hacking if evaluated carelessly, so we also address out-of-sample tracking and IS-OOS degradation in this study. In Sec. \ref{sec:discussion}, we provide conclusive arguments why the success of our framework is real research and cannot be attributed to p-hacking. Overall, these results strongly suggest that LLM-driven agentic and evolutionary search is already a promising tool for quantitative research. As frontier models, evaluation harnesses, and market simulators continue to improve, systems of this kind are likely to become an increasingly powerful component of the trading-research workflow.
\section*{Acknowledgements}
The authors are grateful for the multiple discussions they had with William Cottrell, who provided valuable feedback throughout the time period this research was conducted. Additionally, the authors acknowledge the support of Google and Anthropic, who provided free access to their LLM products via API credits. 
\bibliographystyle{plainnat}
\bibliography{references}

\appendix

\section{Detailed Trading Simulation Setup}
\label{app:simulation_details}

In this appendix we explain our trading simulation setup in enough detail to make our results reproducible. Additional explanations are provided in Sec. \ref{sec:tradingsim}.

\subsection{Data and Splits}
For all experiments, we use BTCUSD 1-minute historical bar data from \texttt{polygon}\footnote{https://massive.com/docs/rest/crypto/overview}, including open (O), high (H), low (L), close (C), and volume (V). We use chronological splits:
\begin{itemize}
    \item \textbf{Train}: 2022-01-01 $\rightarrow$ 2023-12-31 (alpha model training only)
    \item \textbf{Validation}: 2024-01-01 $\rightarrow$ 2024-12-31 (strategy evolution / IS)
    \item \textbf{Test}: 2025-01-01 $\rightarrow$ 2025-10-10 (OOS evaluation)
\end{itemize}

\subsection{Exchange Response Model}
We model a purely passive execution strategy with a 1-minute decision interval. At each minute, the strategy outputs a signed submitted BTC quantity $\Delta q_t^{\mathrm{ord}}$ and a limit price $p_t^{\text{limit}}$.

\subsubsection{Fill Logic}
At each interval, the simulator determines whether the resting order fills based on the observed candle range:
\begin{itemize}
    \item \textbf{Buy} fills if $L < p_t^{\text{limit}}$
    \item \textbf{Sell} fills if $H > p_t^{\text{limit}}$
\end{itemize}
If a fill occurs, execution is at the limit price (no additional slippage). The executed quantity is
\[
\Delta q_t^{\mathrm{fill}} = \Delta q_t^{\mathrm{ord}} \cdot \text{hit\_ratio},
\]
where $\text{hit\_ratio} \in [0,1]$ controls partial fills (default $\text{hit\_ratio}=1$). If neither condition is met, the realized trade quantity is $\Delta q_t^{\mathrm{fill}}=0$.

\subsubsection{Order Lifecycle}
Within each interval:
\begin{enumerate}
    \item \texttt{exchange\_response()} checks whether the prior order filled.
    \item Portfolio position is updated: $\text{position} \leftarrow \text{position} + \Delta q_t^{\mathrm{fill}}$.
    \item \texttt{cancel\_open\_orders()} removes resting orders.
    \item \texttt{set\_passive\_order\_data()} computes new trade quantity and limit price.
    \item \texttt{submit\_order()} submits if the limit price is valid (not \texttt{NaN}).
    \item \texttt{log()} records state.
\end{enumerate}
Only one order is active at any time; each interval replaces the previous order.

\subsection{PnL Calculation}

Let $q_t$ be the BTC quantity held at the start of interval $t$, $\Delta q_t^{\mathrm{fill}}=q_{t+1}-q_t$ be the realized filled BTC trade, $m_t$ be mid price, and $\delta m_t = m_{t+1}-m_t$. Lowercase $q$ quantities are always BTC-denominated in this subsection. We record:
\begin{itemize}
    \item $PnL_t^{\text{pos}} = q_{t+1}\delta m_t$ (frictionless position PnL)
    \item $PnL_t^{\text{target}} = q_{t+1}^{\text{target}}\delta m_t$ (target PnL)
    \item $PnL_t = q_{t+1}\delta m_t - (p_t^{\text{limit}}-m_t)\Delta q_t^{\mathrm{fill}} - 0.00015m_t|\Delta q_t^{\mathrm{fill}}|$ (net PnL)
    \item $PnL_t^{\text{adj}} = PnL_t - I_t$ (impact-adjusted PnL, primary optimization metric), where $I_t$ is the sum of the trade-level impact costs from the market-impact model below for trades executed during interval $t$.
\end{itemize}

\subsection{Market Impact Model}
\label{ssec:market_impact}
To estimate execution costs beyond explicit exchange fees, we use a square-root impact model with power-law temporal decay, belonging to the propagator family of market-impact models \citep{bouchaud2009impact}. The model captures two empirical regularities: price impact scales as the square root of trade size relative to market volume, and a transient component decays over time as the order book recovers. The parameters are calibrated to Hyperliquid BTC-USD perpetuals and are applied post-hoc to backtest trade logs.

Let $\{(Q_i,t_i)\}_{i=1}^N$ be the sequence of executed trades used by the impact model, where $Q_i$ is the signed trade size in USD notional and $t_i$ is the execution timestamp. For an execution during interval $t$ at execution price $p_t^{\mathrm{exec}}=p_t^{\mathrm{limit}}$, this notional is
\begin{equation}
    Q_i = p_t^{\mathrm{exec}}\Delta q_t^{\mathrm{fill}}.
\end{equation}
Thus $Q_i>0$ for buys, $Q_i<0$ for sells, and $|Q_i|$ is the dollar notional traded. The fractional price displacement at time $t$ due to the cumulative effect of all trades up to and including $t$ is
\begin{equation}
    D(t) = \alpha_{\mathrm{perm}} \sum_{i:\, t_i \le t} s_i
         + \alpha_{\mathrm{trans}} \sum_{i:\, t_i \le t} s_i G(t-t_i),
\end{equation}
where $s_i$ is the signed size-impact factor
\begin{equation}
    s_i = \operatorname{sign}(Q_i)\left(\frac{|Q_i|}{V}\right)^\delta,
\end{equation}
and $G(\tau)$ is the temporal decay kernel
\begin{equation}
    G(\tau) = \left(\frac{\tau_0}{\tau+\tau_0}\right)^\beta.
\end{equation}
The kernel satisfies $G(0)=1$ and $G(\tau)\to0$ as $\tau\to\infty$, with asymptotic power-law decay $G(\tau)\sim\tau^{-\beta}$ for $\tau\gg\tau_0$.

The execution cost attributed to trade $i$ is the displacement at the time of execution multiplied by the signed trade size,
\begin{equation}
    c_i = D(t_i)Q_i.
\end{equation}
The interval-level impact charge used in $PnL_t^{\mathrm{adj}}$ is
\begin{equation}
    I_t = \sum_{i:\, t_i \in [t,t+1)} c_i,
\end{equation}
and the total impact cost over the full backtest is
\begin{equation}
    C = \sum_{i=1}^{N} c_i.
\end{equation}
We also report the corresponding cost in basis points of traded notional as
\begin{equation}
    C_{\mathrm{bps}} = \frac{C}{\sum_{i=1}^{N}|Q_i|}\times 10{,}000.
\end{equation}
By construction, $D(t_i)$ includes the self-impact of trade $i$ because the sums include $t_i$ and $G(0)=1$. This charges the full self-impact rather than the average execution price impact, which would be approximately half the self-impact for concave impact functions; the resulting overestimate is deliberate and conservative.

The market-impact parameters are listed in Table~\ref{tab:market_impact_params}.
\begin{table}[h]
\centering
\begin{tabular}{lll}
\toprule
\textbf{Parameter} & \textbf{Symbol} & \textbf{Default} \\
\midrule
Daily market volume & $V$ & \$2B \\
Permanent impact coefficient & $\alpha_{\mathrm{perm}}$ & 0.005 \\
Transient impact coefficient & $\alpha_{\mathrm{trans}}$ & 0.010 \\
Characteristic decay time & $\tau_0$ & 300 s  \\
Decay exponent & $\beta$ & 0.5 \\
Size exponent & $\delta$ & 0.5 \\
\bottomrule
\end{tabular}
\caption{Market impact model parameters}
\label{tab:market_impact_params}
\end{table}

\section{Skeleton Code for the Base Strategy and Forecaster}
\label{app:skeleton_code}

We summarize our baseline forecaster and trading algorithm, which are simplified versions of the basis for the algorithm evolution in this paper. Additional explanations are provided in Sec. \ref{sec:basestratexplain}.

\subsection{Baseline Alpha Forecaster}
\label{ssec:baseline_alpha}

OHLCV bars are transformed into three exponential-moving-average (EMA) return features, and a Ridge regressor is trained to predict future returns over 1, 10, 100, and 1000 minutes.

\begin{lstlisting}[style=compactpython]
import pandas as pd
from sklearn.linear_model import Ridge

def default_calcset(ohlcv: pd.DataFrame) -> pd.DataFrame:
    """OHLCV -> feature matrix."""
    close = ohlcv["close"]
    returns = close.pct_change()
    return pd.DataFrame({
        "ema_ret_1": returns.ewm(span=1).mean(),
        "ema_ret_5": returns.ewm(span=5).mean(),
        "ema_ret_10": returns.ewm(span=10).mean(),
    }).fillna(0)

def train_forecaster(ohlcv: pd.DataFrame):
    """Fit Ridge on EMA features to predict multi-horizon future returns."""
    X = default_calcset(ohlcv).to_numpy()

    y = pd.DataFrame({
        "return_1m": ohlcv["close"].pct_change(1).shift(-1),
        "return_10m": ohlcv["close"].pct_change(10).shift(-10),
        "return_100m": ohlcv["close"].pct_change(100).shift(-100),
        "return_1000m": ohlcv["close"].pct_change(1000).shift(-1000),
    }).fillna(0)

    model = Ridge(alpha=0.5)
    model.fit(X, y.to_numpy())
    return model

def forecast_alpha(ohlcv: pd.DataFrame, model):
    """Fresh OHLCV -> features -> alpha predictions at each horizon."""
    X = default_calcset(ohlcv).to_numpy()
    return model.predict(X)
\end{lstlisting}

\subsection{Base Strategy}
\label{app:basestrat}

The alpha forecast is mapped to a target position, and the strategy then submits a passive buy or sell limit order to move inventory toward that target. 

\begin{lstlisting}[style=compactpython]
import numpy as np

BUY = "B"
SELL = "A"

class DefaultPassiveExecutor:
    def __init__(
        self,
        sizing_factor=10_000,
        q_max=200_000,
        max_trade_frac=0.2,
        min_trade_size_usd=0,
        alpha_adjustment_knob=0.5,
        risk_reduction_factor=0.6,
        zp=0.0001,
        zp_riskoff=0.00003,
        fast_flat_minutes=10,
        std=1,
    ):
        self.sizing_factor = sizing_factor
        self.q_max = q_max
        self.max_trade_frac = max_trade_frac
        self.min_trade_size_usd = min_trade_size_usd
        self.alpha_adjustment_knob = alpha_adjustment_knob
        self.risk_reduction_factor = risk_reduction_factor
        self.zp = zp
        self.zp_riskoff = zp_riskoff
        self.fast_flat_minutes = fast_flat_minutes
        self.std = std
        self.context_correction_factor = 0

    # EVOLVE-BLOCK-TARGET-START
    def set_target(self, state):
        """Alpha + market state -> signed target trade side."""
        alpha = state["alpha"]
        alpha_sd = state["alpha_sd"]
        mid = state["mid"]
        mid_book = state["mid_book"]
        q_x = state["position_btc"]
        data_lag_minutes = state.get("data_lag_minutes", 0)

        limit_order_depth = self.std * self.zp
        expected_fee = max(0.015 / 100 - limit_order_depth, 0.005 / 100)

        realized_alpha = np.log(mid_book / mid)
        alpha_corrected = alpha - self.context_correction_factor * realized_alpha
        q_usd = np.nan_to_num(q_x * mid_book)

        k = self.sizing_factor / alpha_sd
        small_alpha = abs(alpha_corrected - q_usd / k) < expected_fee
        wrong_direction = np.sign(q_x * alpha_corrected) < 0
        risk_reduction_mode = small_alpha and wrong_direction

        if risk_reduction_mode:
            target_pos_usd = q_usd * self.risk_reduction_factor
        elif (
            abs(realized_alpha) * self.context_correction_factor > abs(alpha)
            and np.sign(realized_alpha * alpha) > 0
        ):
            target_pos_usd = q_usd
        else:
            long_target_usd = self.sizing_factor * (alpha_corrected - expected_fee) / alpha_sd
            short_target_usd = self.sizing_factor * (alpha_corrected + expected_fee) / alpha_sd

            if long_target_usd > q_usd:
                target_pos_usd = long_target_usd
            elif short_target_usd < q_usd:
                target_pos_usd = short_target_usd
            else:
                target_pos_usd = q_usd

        lag_adjustment = 1 - min(data_lag_minutes, self.fast_flat_minutes) / self.fast_flat_minutes
        if risk_reduction_mode:
            correction_factor = lag_adjustment
        else:
            risk_adjustment = 1 - np.tanh(abs(q_usd) / self.q_max) * self.alpha_adjustment_knob
            correction_factor = risk_adjustment * lag_adjustment

        target_pos_usd = np.clip(target_pos_usd * correction_factor, -self.q_max, self.q_max)
        target_pos_btc = target_pos_usd / mid_book

        raw_trade_qty = target_pos_btc - q_x
        max_trade_btc = self.max_trade_frac * self.q_max / mid_book
        target_trade_qty = np.clip(raw_trade_qty, -max_trade_btc, max_trade_btc)

        delta_usd = abs(target_pos_btc - q_x) * mid_book
        if target_pos_btc > q_x and delta_usd > self.min_trade_size_usd:
            side = BUY
        elif target_pos_btc < q_x and delta_usd > self.min_trade_size_usd:
            side = SELL
        else:
            side = None

        return {
            "side": side,
            "target_trade_qty": target_trade_qty,
            "risk_reduction_mode": risk_reduction_mode,
            "limit_order_depth": limit_order_depth,
        }
    # EVOLVE-BLOCK-TARGET-END

    # EVOLVE-BLOCK-LIMIT-START
    def set_limit_order(self, state, target):
        """Signed target trade -> passive limit order."""
        if target["side"] is None:
            return None

        mid_book = state["mid_book"]
        side_multiplier = np.sign(target["target_trade_qty"])

        if target["risk_reduction_mode"]:
            depth = self.zp_riskoff * self.std
        else:
            depth = target["limit_order_depth"]

        limit_price = mid_book * np.exp(-side_multiplier * depth)
        return {
            "side": target["side"],
            "limit_price": limit_price,
            "target_trade_qty": target["target_trade_qty"],
        }
    # EVOLVE-BLOCK-LIMIT-END

    def set_passive_order_data(self, state):
        target = self.set_target(state)
        return self.set_limit_order(state, target)

def apply_order_constraints(order, mid_book, max_limit_order_usd=100_000):
    """Non-evolvable post-processing applied after the strategy creates an order."""
    if order is None:
        return None

    if order["side"] == BUY and order["target_trade_qty"] < 0:
        order["target_trade_qty"] = abs(order["target_trade_qty"])
    elif order["side"] == SELL and order["target_trade_qty"] > 0:
        order["target_trade_qty"] = -abs(order["target_trade_qty"])

    order_usd = abs(order["target_trade_qty"] * mid_book)
    if order_usd > max_limit_order_usd:
        max_btc = max_limit_order_usd / mid_book
        order["target_trade_qty"] = max_btc * np.sign(order["target_trade_qty"])

    return order
\end{lstlisting}

\section{Human-Edited Summaries of AI-Generated Run Reports}
\label{app:run_report_summaries}

This appendix collects qualitative summaries of the most salient modifications introduced in Runs~1--5. The text in the shaded boxes is not raw model output: it is human-edited material distilled from the AI-generated run reports produced during the corresponding \textsc{MadEvolve} experiments. We keep these summaries in the appendix so that the main text can focus on experimental setup, quantitative outcomes, and interpretation.

\subsection{Run 1: Target Position Evolution}

\begin{airptbox}
\paragraph{Structural modifications.}
The evolved strategy introduces several qualitative changes to the target computation that go well beyond parameter tuning. The most significant is the introduction of \emph{stateful alpha processing}. While the baseline operates statelessly on the raw corrected alpha at each time step, the evolved variant maintains exponential moving averages of the alpha signal, alpha dispersion, and mid-price movement. A momentum term, defined as the deviation of the instantaneous alpha from its EMA, amplifies signals that are building in conviction and suppresses transient spikes. This momentum is further dampened when a composite measure of alpha dispersion and micro-volatility is elevated, preventing the strategy from chasing noisy signals in choppy conditions.

The second major innovation is a \emph{dynamic hysteresis band} that replaces the baseline's static fee-based entry threshold. The evolved strategy computes an impact-aware no-trade band that widens with inventory utilization and a churn proxy derived from recent turnover. A trade is initiated only when the signal urgency, measured as the absolute difference between the scaled alpha and an inventory-implied alpha, exceeds this band. The inventory-implied alpha translates the current position into signal space, providing a natural reference against which new signals are compared.

Third, the evolved strategy introduces \emph{urgency-based execution sizing}. Rather than trading the full target delta in a single step, the strategy computes a step fraction as a concave function of how far the signal extends beyond the no-trade band. This fractional execution, subject to a minimum step size of 2.4\% of the target, allows the strategy to scale into positions gradually when conviction is moderate and act more decisively under strong signals.

\paragraph{Signal transformation.}
The corrected alpha undergoes a nonlinear transformation before entering the position calculation. After blending with the momentum component and subtracting the realized price drift, the signal is normalized by alpha volatility and passed through a signed power function with exponent slightly above unity ($\approx 1.07$). This mild convexity amplifies strong signals relative to weak ones, effectively suppressing noise-driven trades while preserving responsiveness to genuine opportunities.

\paragraph{Risk management.}
Risk management in the evolved strategy operates through a hyperbolic tangent inventory penalty with a raised exponent ($\approx 1.55$), which delays the onset of position compression at moderate utilization but applies aggressive dampening as inventory approaches capacity. A dedicated risk reduction mode activates when the position opposes the alpha signal and the signal lacks sufficient urgency to overcome the no-trade band; in this mode, the target is scaled to approximately 67\% of the current position, forcing a gradual unwind.
\end{airptbox}

\subsection{Run 2: Order Placement Evolution}

\begin{airptbox}
\paragraph{Optimization-based depth selection.}
The most striking innovation is the replacement of the baseline's rule-based depth calculation with an explicit optimization procedure. The evolved strategy defines an expected utility objective over quoting depth,
\begin{equation}
    U(d) = \hat{e}(d) \cdot \hat{p}_{\text{fill}}(d) - \lambda_{\text{tox}} \cdot R_{\text{tox}}(d) - \lambda_{\text{imp}} \cdot R_{\text{imp}}(d),
\end{equation}
where $\hat{e}(d)$ is the expected edge at depth $d$, $\hat{p}_{\text{fill}}(d)$ is an estimated fill probability that decreases with depth, $R_{\text{tox}}(d)$ is a toxicity risk term that grows superlinearly as depth shrinks, and $R_{\text{imp}}(d)$ penalizes market impact proportional to order size divided by depth. A deterministic ternary search over a bounded interval finds the depth maximizing this objective at each time step. The fill probability is modeled as a sigmoid function of alpha alignment, a composite safety score, and normalized depth, with a tunable sensitivity parameter governing how rapidly fill probability decays deeper into the book.

\paragraph{Fair value anchoring.}
Rather than computing the limit price as an offset from the raw mid-book price, the evolved strategy first shifts the reference to a fair value mid-price,
\[
    m_{\text{fair}} = m \cdot \exp(\omega_{\alpha} \cdot \alpha),
\]
where $\omega_{\alpha} \approx 0.6$ controls how aggressively pricing tracks the alpha prediction. This anchoring effectively skews quoting in the direction of the predicted move, tightening the spread on the side supported by the signal and widening it on the opposing side.

\paragraph{Microstructure state tracking.}
The evolved strategy introduces stateful tracking of three market microstructure features via exponential moving averages: mid-price momentum, alpha jump magnitude, and realized pickoff rate. These features feed into a composite toxicity score that dynamically adjusts both quoting depth and order quantity. When the mid EMA shows strong directional momentum, or when a large alpha jump is detected, the strategy widens its quotes to reduce adverse selection exposure. Realized pickoff events, measured as adverse mid-price movements following a fill, receive particularly high weight ($\approx 7.8$) in the toxicity score, reflecting the finding that adverse selection on recently filled orders is the dominant source of execution cost.

\paragraph{Dynamic quantity shaping.}
The baseline applies a static multiplier to the target trade quantity. The evolved strategy instead computes a dynamic quantity as the product of several factors: a base multiplier ($\approx 1.28$), a conviction boost that scales quadratically with alpha alignment, a risk-reducing boost that increases order size when the trade would flatten inventory, and a dampening factor derived from the composite toxicity and fill confidence scores. This shaping ensures that the strategy trades large when conditions are favorable and retreats to minimal sizes when adverse selection risk is elevated.

\paragraph{Asymmetric alpha response.}
The strategy differentiates its quoting behavior depending on whether the alpha signal supports or opposes the order side. When alpha aligns with the order direction, the depth adjustment tightens the quote ($\omega_{\text{urgency}} \approx 0.9$), seeking a faster fill at a price closer to fair value. When alpha opposes the order, the depth is widened ($\omega_{\text{defense}} \approx 0.43$), reflecting the higher cost of being adversely selected on the wrong side.
\end{airptbox}

\subsection{Run 3: Joint Target and Order Evolution}

\begin{airptbox}
\paragraph{Joint architecture.}
Unlike Runs~1 and~2, where one component was fixed at its baseline, Run~3 permits the evolution to restructure the interaction between signal processing, position sizing, cost modeling, and order pricing. The resulting strategy exhibits tight integration between these components: the same alpha state feeds the no-trade band, the target sizing, and the reservation price; the inventory ratio modulates effective alpha, target leverage, trade cap, and quote depth; and execution mode (normal vs.\ risk-reducing) gates a separate quoting branch.

\paragraph{Multi-horizon alpha smoothing and exhaustion correction.}
The evolved strategy maintains three exponential moving averages over the alpha stream---a fast EMA at decay rate $0.20$, a slow EMA at $0.05$, and an alpha-volatility EMA at $0.10$ that tracks $|\alpha - \overline{\alpha}_{\text{fast}}|$. These produce a soft consistency gate: when the instantaneous alpha and the fast EMA disagree in sign, the corrected alpha is downweighted to a quarter of its raw value. On top of the baseline stale-information correction $\alpha - c \cdot \log(\text{mid\_book}/\text{mid})$, the strategy adds an explicit \emph{signal-exhaustion} term: if the realised mid move since the last alpha publication has already gone in the alpha direction, the corrected alpha is shrunk by up to 95\% in proportion to how much of the original signal magnitude has already been spent. The two corrections together prevent the strategy from chasing moves that have largely already happened, a failure mode of the baseline at minute frequency.

\paragraph{Inventory-dependent alpha penalty and dynamic no-trade band.}
The strategy folds inventory pressure into the effective alpha rather than the target itself: $\alpha_{\text{eff}} = \alpha_{\text{corr}} - 1.8 \, q_r \, \alpha_{\text{sd}} \, (1 + 1.5\,|q_r|)$, where $q_r \in [-1,1]$ is the normalised position ratio. The penalty is approximately linear at low $|q_r|$ and superlinear---roughly cubic in displacement---near the position limit, creating a soft mean-reversion barrier without the hard cliff of an explicit position cap. A dynamic no-trade band $\text{deadband} = \text{fee} \cdot (1.2 + 0.5 \cdot \alpha_{\text{vol}}/\alpha_{\text{sd}})$ then suppresses trades whose effective alpha is too small to clear the fee net of a volatility-aware buffer; the buffer widens when the realised dispersion of $\alpha$ exceeds its standard estimate, so the strategy becomes more conservative in regimes where the signal itself is unstable. The band is overridden only when the signal points against an already-large inventory, allowing risk-reducing trades to fire at smaller magnitudes than risk-adding trades. A separate hard cutoff rejects any signal-aligned trade once $|q_r| > 0.98$.

\paragraph{Saturating sizing with hysteresis.}
The raw target position is computed as $\text{target} = q_{\max} \cdot \tanh(s \cdot \ell)$, where $s = \alpha_{\text{eff}}/\alpha_{\text{sd}}$ is signal conviction and $\ell = 1.3 - 0.3 \, |q_r|$ is an inventory-tapered leverage. The $\tanh$ function gives the strategy near-linear response at modest conviction but saturates well below $q_{\max}$ once the signal is strong, replacing the baseline's hard clip with a smooth ceiling. Around this target the strategy imposes a hysteresis band of $2.5\%$--$4\%$ of $q_{\max}$ (tighter when conviction is high): the target only updates when the new value differs from the cached one by more than the band, which suppresses high-frequency churn driven by noisy minute-to-minute alpha fluctuations.

\paragraph{Turnover governor and conviction-aware quoting.}
A turnover governor tracks an EMA of per-minute notional traded; once that EMA exceeds $15\%$ of $q_{\max}$, the next non-risk-reducing trade is halved, throttling the strategy in regimes where it would otherwise rack up impact cost faster than the signal repays. Trade size itself is scaled by a conviction boost $1 + 0.7 \, |s|^{1.2}$, with an additional $1.2\times$ multiplier in the risk-reducing branch so the strategy can flatten faster than it accumulates. Quote depth is asymmetric: in the normal branch, $\text{depth}_{\text{mult}} = 1.15 - 0.7 \, \tanh(|s|)$ tightens the quote toward mid as conviction rises, multiplied by a skew $\text{clip}(1 + 1.4 \, q_r \, \text{side}, 0.15, 2.8)$ that pulls quotes away from mid when the trade would extend an already-large inventory and toward mid when it would reduce it. The risk-reducing branch instead uses the dedicated \texttt{zp\_riskoff} parameter scaled by an urgency factor $\text{clip}(|s|/0.3, 0.5, 2.5)$, so unwinding a losing position quotes more aggressively the stronger the contrary signal becomes.
\end{airptbox}

\subsection{Run 4: Feature Evolution for Alpha Prediction}

\begin{airptbox}
\paragraph{Multi-scale band-pass momentum.}
Where the baseline computes three demeaned EMAs of one-step returns at a single demeaning window per feature, the evolved \texttt{default\_calcset} structures its momentum features as three band-pass tiers, each tuned to a distinct timescale relative to the 10-minute prediction target: short (\texttt{hls=[2,5]}, \texttt{demean\_hl=20}), medium (\texttt{hls=[10,20]}, \texttt{demean\_hl=60}), and long (\texttt{hls=[60]}, \texttt{demean\_hl=240}). The progression of demeaning windows produces a coarse band-pass decomposition of the return series in which both short-lived noise and slowly-drifting trend components are attenuated, leaving the predictable mid-frequency component to dominate the input to the ridge regression.

\paragraph{Mean-reversion state and order-flow features.}
The strategy adds a 60-minute deviation from a local mean, $z_{60} = \text{mid} - \text{EMA}(\text{mid}, 60)$, together with its velocity computed as the EMA difference of $z_{60}$ at halflives 5 versus 20. The pair captures both the current displacement of the mid-price from its short-run mean and the rate at which that displacement is being unwound, providing the predictor with explicit mean-reversion state. On the volume side, three orthogonal order-flow features are introduced. A volume surprise term, $\text{EMA}(\log V, 3) - \text{EMA}(\log V, 60)$, distinguishes short bursts of activity from the slowly-varying baseline. A signed volume flow term, $\text{EMA}(r_t \cdot \log V_t, 12)$, weighs each return increment by the contemporaneous log volume to approximate the directional component of order flow. Finally an acceleration term, the EMA difference of the signed flow at halflives 5 versus 20, exposes regime shifts in order-flow direction. None of these have analogues in the baseline, which is purely price-based.

\paragraph{Stability post-processing.}
After feature construction, every column is winsorized globally at $\pm 5\sigma$. The motivation given in the evolved code's docstring is to ``stabilize ICIR''; operationally, the clipping prevents heavy-tailed events from inflating the standard deviation of the daily ICs, which is the denominator of the ICIR component of Equation~\ref{eq:run4_score} and which the LLM evidently identified as the binding term in the fitness function.
\end{airptbox}

\subsection{Run 5: Joint Feature and Strategy Evolution}

\begin{airptbox}
\paragraph{Evolved feature pipeline.}
The evolved \texttt{default\_calcset} extends the baseline three-feature momentum basis into a markedly richer representation that combines multi-scale momentum, volatility-regime context, volume-driven order-flow proxies, and nonlinear interaction terms. At the momentum core, the baseline call to \texttt{construct\_return\_calcs} is broadened to halflives drawn from a Fibonacci-like ladder, \texttt{hls=[1,2,3,5,8,13,21,34,55,89]}, with a single demeaning halflife of 66~minutes that brackets the 10-minute prediction horizon; Bollinger pressure at four horizons, a simple RSI at three horizons, and a mid-minus-mean deviation at 66~minutes are retained from the library. On top of this core the evolved code introduces four new families. A \emph{volatility regime} group defines $\text{vol}_\text{fast}/\text{vol}_\text{slow}$ from EMA returns at halflives 5 and 66, together with a squared-deviation term $(\text{vol\_ratio}-1)^2$ and a volatility acceleration computed as the log return of $\text{vol}_\text{fast}$; the squared term gives the predictor an explicit ``how unusual is current volatility'' signal, which is sign-agnostic by construction. A \emph{range-based stress} group derives a $\log(H/L)$ range proxy, normalises it by the slow volatility scale, and then computes its deviation from a 30-minute EMA, giving a short-window stress indicator that is not expressible as a pure EMA of minute returns. A \emph{volume/order-flow} group introduces a log-volume impulse $\log(V/\text{EMA}(V,45))$, a signed flow term $r_t \cdot V_t / \text{EMA}(V,45)$ smoothed at fast (10) and slow (45) halflives, and a flow-divergence term equal to their difference; these quantities have no analogue in the baseline, which is purely price-based. Finally, a \emph{trend-quality and interaction} group combines a dimensionless trend strength $\text{ema\_diff}(\text{mid},10,45)/\text{vol}_\text{slow}$ at two scales (and their geometric-signed product \texttt{trend\_align}), an efficiency-ratio trend-quality term $|\text{EMA}(r,20)|/\text{EMA}(|r|,20)$, and three multiplicative interactions (trend$\times$volatility, flow$\times$trend, volume impulse$\times$trend) that let the ridge predictor pick up regime-conditional momentum. Every new feature is aggressively clipped---volatility ratios to $[0.2,5]$, log-volume impulses to $[-4,4]$, flows to $\pm10^{-3}$---which plays the same role as the $\pm5\sigma$ winsorisation in Run~4: it prevents a handful of extreme events from driving the ridge solution or inflating the dispersion of the daily IC.

\paragraph{Evolved execution strategy.}
The evolved \texttt{set\_passive\_order\_data} is structurally more integrated than the baseline and closely mirrors, in spirit, the innovations discovered independently in Runs~1 and~2. The fee estimate is shifted upward by a calibrated impact buffer of 4.5~bps and scaled by a factor $(1 + 0.22 \cdot |q|/q_{\max})$ that makes the effective cost grow with inventory utilisation, so that the no-trade band widens as the book fills up. Alpha is corrected for realised price drift via a \texttt{context\_correction\_factor} coefficient applied to $\log(\text{mid\_book}/\text{mid})$, matching the stale-information correction used by the baseline but with an evolvable mixing weight. Target sizing introduces a \emph{super-linear conviction boost} of the form $1 + 3.0 \cdot |\alpha_z|^{1.25}$ (with $\alpha_z$ capped at 3.8), applied to an effective sizing factor; the combination implements a soft power-law response that is nearly identity near zero alpha but amplifies high-conviction signals by roughly $5\times$ at the cap. Inventory risk is handled by a parabolic-plus penalty, $1 - (|q|/q_{\max})^{2.2} \cdot \kappa$, that keeps the strategy close to its unconstrained target at low utilisation and dampens aggressively only near the position cap. A separate risk-reduction branch activates when a small alpha is pointing against the current position, scaling the target to \texttt{risk\_reduction\_factor}$\cdot q$. Trade-size limits mirror the target logic: the per-minute trade cap is inflated by a second conviction boost $1 + 3.8 \cdot |\alpha_z|^{1.2}$ (with $\alpha_z$ capped at 3.2), so the strategy can close a larger fraction of the target-current gap in a single minute when the signal is strong. Finally, the limit-order depth is written as a volatility-scaled base depth multiplied by $(1.18 - 0.88 \cdot \min(|\alpha_z|,2))$, which tightens the quote toward the mid as the signal strengthens, and is widened by 35\% in the risk-off branch. The overall pattern is that every sizing and pricing decision is monotone in signal conviction $|\alpha_z|$, and monotone in inventory utilisation $|q|/q_{\max}$, with the two effects composed multiplicatively rather than additively.
\end{airptbox}

\section{Hyperparameter Calibration Details}
\label{app:hyperparam_details}

The eight quantities in Table~\ref{tab:hyperparam_params} are the parameters used by the baseline passive-execution strategy in Appendix~\ref{app:basestrat}. Here, we report the search bounds and the values selected by Optuna in the two calibration sweeps.

\begin{table}[htbp]
\centering
\footnotesize
\setlength{\tabcolsep}{2pt}
\renewcommand{\arraystretch}{1.04}
\begin{tabular}{p{0.17\textwidth}p{0.29\textwidth}ccp{0.08\textwidth}@{\hspace{2pt}}cc}
\toprule
\textbf{Parameter} & \textbf{Description} & \textbf{Default} & \textbf{Bounds} & \textbf{Scale} & \shortstack{\textbf{Optimal}\\\textbf{(Baseline)}} & \shortstack{\textbf{Optimal}\\\textbf{(Evolved)}} \\
\midrule
\path{sizing_factor} & How strongly the forecast is converted into desired dollar exposure; higher values trade larger positions for the same signal. & 10{,}000 & 500--50{,}000 & log & 49{,}816 & 26{,}436 \\
\path{max_trade_frac} & Largest order allowed as a fraction of the position limit; prevents building or unwinding too much in one minute. & 0.20 & 0.01--0.5 & linear & 0.493 & 0.488 \\
\path{min_trade_size_usd} & Smallest desired position change worth sending as an order; smaller adjustments are skipped. & 0 & 0--5{,}000 & linear & 751 & 382 \\
\path{alpha_adjustment_knob} & Strength of the inventory brake in normal trading; higher values reduce targets more when inventory is large. & 0.50 & 0--1 & linear & 0.431 & 0.071 \\
\path{risk_reduction_factor} & Fraction of the current position retained when the signal is weak but points against the existing position. & 0.60 & 0--1 & linear & 0.355 & 0.295 \\
\path{zp} & Limit price offset knob for normal orders; smaller values post closer to the mid price and fill more easily. & $10^{-4}$ & $10^{-6}$--$10^{-2}$ & log & $1.8 \times 10^{-3}$ & $1.0 \times 10^{-6}$ \\
\path{zp_riskoff} & Same limit price offset knob, but for risk-reduction orders; smaller values unwind closer to the mid price. & $3.0 \times 10^{-5}$ & $10^{-6}$--$10^{-2}$ & log & $1.3 \times 10^{-4}$ & $1.1 \times 10^{-3}$ \\
\path{fast_flat_minutes} & How quickly stale data pushes the target toward zero; after this many lagged minutes the target is zero. & 10 & 2--60 & linear & 57.9 & 38.4 \\
\bottomrule
\end{tabular}
\caption{Effective hyperparameters optimized by Optuna in the two calibration sweeps.}
\label{tab:hyperparam_params}
\end{table}

\end{document}